\documentstyle[12pt,a41,rotating,amstex,cite,portland]{article}
\newcommand{\lifhalf}{\Li_4\left(\frac{1}{2}\right)}
\newcommand{\liFhalf}{\Li_5\left(\frac{1}{2}\right)}
\newcommand{\ZTW}{\zeta_2}
\newcommand{\ZTH}{\zeta_3}

\newcommand{\ZFI}{\zeta_5}
\newcommand{\Li}{{\rm Li}}
\newcommand{\Sf}{{\rm S}_{1,2}}
\newcommand{\si}{{\rm sign}}
\newcommand{\ds}{\displaystyle}

\newcommand{\bq}{\begin{equation}}
\newcommand{\eq}{\end{equation}}
\newcommand\beq{\begin{equation}}
\newcommand\eeq{\end{equation}}
\newcommand\bea{\begin{eqnarray}}
\newcommand\eea{\end{eqnarray}}

\newcommand\Mvec{\,\mbox{\bf M}}
\newcommand\Nvec{\,\mbox{\bf N}}
\newcommand\Cvec{\,\mbox{\bf C}}
\newcommand\Rvec{\,\mbox{\bf R}}

\begin{document}
\setlength{\baselineskip}{0.515cm}
\sloppy
\thispagestyle{empty}
\begin{flushleft}
DESY 07--042
\hfill {\tt arXiv:0901.3106 [hep-ph]}
\\
SFB/CPP-09-05\\
January 2009\\
\end{flushleft}

\mbox{}
\vspace*{\fill}
\begin{center}

{\LARGE\bf Structural Relations of Harmonic Sums and}\\

\vspace*{3mm}
{\LARGE\bf \boldmath Mellin Transforms up to Weight  $w = 5$}\\

\vspace{4cm}
\large
Johannes Bl\"umlein

\vspace{1.5cm}
\normalsize
{\it  Deutsches Elektronen--Synchrotron, DESY,}\\
{\it  Platanenallee 6, D-15735 Zeuthen, Germany}
\\

\end{center}
\normalsize
\vspace{\fill}
\begin{abstract}
\noindent
We derive the structural relations between the Mellin transforms of weighted 
Nielsen integrals emerging in the calculation of massless or massive single--scale 
quantities in QED and QCD, such as anomalous dimensions and Wilson coefficients, 
and other hard scattering cross sections depending on a single scale. The set 
of all multiple harmonic sums up to weight five cover the sums needed in the 
calculation of the 3--loop anomalous dimensions. The relations extend the set 
resulting from the quasi-shuffle product between harmonic sums studied earlier. 
Unlike the shuffle relations, they depend on the value of the quantities considered. 
Up to weight {\sf w = 5},~ 242 nested harmonic sums contribute. In the present 
physical applications it is sufficient to consider the sub-set of harmonic sums 
not containing an index $i = -1$, which consists out of 69 sums. The algebraic relations 
reduce this set to 30 sums. Due to the structural relations a final reduction of the 
number of harmonic sums to {15} basic functions is obtained. These functions can be 
represented in terms of factorial series, supplemented by harmonic sums which are 
algebraically reducible. Complete analytic representations are given for these {15} 
meromorphic functions in the complex plane deriving their asymptotic- and recursion 
relations. A general outline is presented on the way nested harmonic sums and multiple 
zeta values emerge in higher order calculations of zero- and single scale quantities.
\end{abstract}

\vspace*{\fill}
\noindent
\numberwithin{equation}{section}
\newpage
\section{Introduction}
%

\vspace{1mm}
\noindent
Finite harmonic sums~\cite{HS1,HS2,HS3} and associated to them, by a {Mellin} 
transform, { Nielsen}--type integrals~\cite{NIELS1}, emerge in perturbative 
calculations of massless and massive single scale problems in Quantum Field-Theory.
The Wilson coefficients and anomalous dimensions in deeply inelastic scattering 
to three loops \cite{ANDCO}, the massive quark Wilson coefficients 
in the limit $Q^2 \gg m^2$ \cite{HEAV,HEAV1}, as well as the Wilson coefficients for 
the Drell-Yan process, hadronic Higgs production in the heavy mass limit, the 
time-like coefficient functions for parton fragmentation into hadrons, 
cf. \cite{TDYHI}, the soft-- and virtual corrections to Bhabha scattering 
\cite{BHABHA}, and various processes more belong to this class. At 2--loop order
the respective expressions in terms of harmonic sums were given in \cite{COHS} 
using the algebraic and structural relations between these quantities of 
weight {\sf w $\leq$ 4}. Due to the complexity of higher 
order calculations the knowledge of as many as possible relations between the 
finite harmonic sums is of importance to simplify the calculations and to obtain
compact analytic results. It turns out that, up to isomorphisms, the 
basic functions, which express the physical quantities mentioned above, are always 
the same and form a basis of an algebra, over which the observables can be 
expressed.
These basic functions therefore form central objects in the 
description of a wide range of physical quantities. 

To derive the basic functions we first consider the multiple finite harmonic sums 
$S_{a_1, \ldots, a_n}(N)$. They provide a unique language to express single scale
quantities, which emerge in field theoretic calculations.
They are defined by
\begin{eqnarray}
\label{eqHS}
S_{a_1, \ldots, a_n}(N) = \sum_{k_1 = 1}^N \sum_{k_2 = 1}^{k_1} \ldots
\sum_{k_n = 1}^{k_{n-1}} \frac{\si(a_1)^{k_1}}{k_1^{|a_1|}} \ldots
\frac{\si(a_n)^{k_n}}{k_n^{|a_n|}}~.
\end{eqnarray}
Here, $a_k$ are positive or negative integers and $N$ is a positive
even or odd number. One calls $n$ the {\sf depth} and $\sum_{k = 1}^n
|a_k|$ the {\sf weight} of a harmonic sum. Harmonic sums are associated to
{ Mellin} transforms of real functions $f(x)$
\begin{eqnarray}
\label{eqMELL}
S_{a_1, \ldots, a_n}(N) = \int_0^1 dx~x^{N}~f_{a_1, \ldots, a_n}(x)
= \Mvec[f_{a_1, \ldots, a_n}(x)](N)~.
\end{eqnarray}
Here $\Mvec$ denotes the Mellin-transform.
The functions 
$f_{a_1, \ldots, a_n}(x)$ are usually
linear combinations of harmonic polylogarithms $H_{b_1, \ldots, b_k}(x)$ 
\cite{POLYL,VR,LILLE},
weighted by the factor $1/(1 \pm x)$. $f(x)$ may be distribution--valued \cite{DIST}.One example for this class is
\begin{equation}
\int_0^1 dx~x^{N}~[f_{a_1, \ldots, a_n}(x)]_+ =  \int_0^1 dx~ \left(x^{N} -1\right) 
{f}_{a_1, \ldots, a_n}(x)~.
\end{equation}

The number of harmonic sums grows exponentially with the weight {\sf w} and
amounts to $3^{{\sf w-1}}$ for all sums of weight $w' \leq w$. 
The finite harmonic sums are not
independent objects 
but obey relations. In view of their rapidly growing number for higher weight
it is worthwhile to reveal all  their relations to express
all harmonic sums in terms of a small set of basic elements only. In a
foregoing paper~\cite{JBSH} we investigated the shuffle relations  
\cite{HOFF} between
general finite alternating or non--alternating harmonic 
sums.~\footnote{Various mathematical terms used in the following were defined 
in Ref.~\cite{JBSH} to which we refer readers from the physics community who 
are not yet used to this terminology.} There all 
shuffle relations for harmonic sums up to depth and weight {\sf w = 6} were derived
in general form and the counting relations for the number of basis
elements spanning the algebra of harmonic sums after the shuffle relations
are used have been given. For a certain class of indices $\{a_1, \ldots,
a_k\}$, where $n_j|_{j=1 \ldots q}$ elements are equal and $n = \sum_{j=1}^q n_j$,
the total number of harmonic sums is
\begin{eqnarray}
\label{eqNUMB}
n(n_1, \ldots, n_q) = 
\frac{n!}
{n_1! \ldots n_q!}~.
\end{eqnarray}
The shuffle relations allow to express these sums in terms of 
\begin{eqnarray}
\label{eqWIT2}
l_n(n_1, \ldots, n_q) = \frac{1}{n} \sum_{d|n_i} \mu(d)
\frac{\left(\frac{n}{d}\right)!}
{\left(\frac{n_1}{d}\right)! \ldots \left(\frac{n_q}{d}\right)!}
\end{eqnarray}
basis elements and polynomials out of objects of lower weight.
Here Eq.~(\ref{eqWIT2}) is the 2nd { Witt} formula, cf.~\cite{WITT}, and   
$\mu(k)$ denotes the { M\"obius} function~\cite{MOB}. $l_n$ also counts
the number of { Lyndon} words~\cite{LYND}  associated to  $\{a_1,
\ldots, a_k\}$, since all harmonic sums belonging to this index set are
{\sf freely} generated by the { Lyndon} words as a consequence of
{ Radford's} theorem~\cite{RADF}.

In the limit $N \rightarrow \infty$ harmonic sums turn into multiple zeta
values (MZVs)~\cite{EZ}~\footnote{If $a_1 =1$, the corresponding multiple sums
diverge for $N \rightarrow \infty$ but can be considered symbolically. Actually they are 
expressible 
in terms of a polynomial of the (divergent) harmonic series $\sigma_0 =
\sum_{k=1}^\infty (1/k)$ and the finite basis elements of multiple zeta
values, see \cite{HS3,JBSH,BBV}.}. Since multiple zeta values obey
many more relations, ~see e.g. \cite{WALD,LISO}, than the shuffle relations the
question arises which other relations may hold for multiple harmonic sums.~\footnote{
The relations between the alternating MZVs 
to weight {\sf w = 12} and non-alternating MZVs 
to weight {\sf w = 24} based on the shuffle-, stuffle-, and new relations
have been determined in \cite{BBV} recently.
Generalizations of finite harmonic sums and harmonic polylogarithms have been 
given for other applications, cf. \cite{GON} and \cite{MUW}.}
A series of relations emerges due to the integral representations of the
harmonic sums which were derived in Ref.~\cite{HS2}. Furthermore, some sums
have a completely symmetric index set and do therefore decompose into polynomials of
single harmonic sums. This also implies relations between the {
Mellin} transforms of certain { Nielsen}--type integrals weighted with the
denominators $1/(1 \pm x)$. 

In the present paper we will not limit the consideration to harmonic sums with 
$N~\in~{\Nvec}$ but will allow for $N~\in~{\Cvec}$ and derive the complex 
analysis for these quantities. Along with this, 
new relations between these objects will emerge. In this generalized 
case we consider rational fractions of $N$,~$N/q$, and allow to differentiate for 
the index  $N$, which both yield new relations. 
The harmonic sums turn out to be meromorphic functions. They can be represented through 
factorial series  \cite{NIELS2,ELAND} up to polynomials of  $S_1^k(N)$
and harmonic sums of lower degree. The analytic continuation of $S_1(N)$ is given by
{ Euler's} di-gamma function $\psi(z)$.

The weight {\sf w = 1} harmonic sums $S_{\pm k}(N)$ may all be traced back to the
$\psi-$function for $N~\in~{\bf C}$. At the beginning of the 20th century 
{ Nielsen}~\cite{NIELS2} studied the respective 
functions belonging to the class of weight {\sf w = 2}, from which the corresponding 
analytic
continuations can be derived. 
Due to the fact, that harmonic sums, with the exception of factors of $S_1(N)$, 
can be represented 
as factorial series, allows to derive their analytic continuations. The recursion relations
for $(N+1) \rightarrow N, N~\in~{\bf C}$, and the asymptotic representation for 
$|N| \rightarrow \infty$ are obtained in analytic form. The singularities are poles located 
at the non--positive integers.
Due to the presence of factors $S_1^k(N)$ in some cases the respective sums behave $\propto \ln^k(N)$
as $|N| \rightarrow \infty$.~\footnote{Precise analytic continuations of the basic functions \cite{HS2,JB04}
based on semi--numerical representations were given in \cite{JB2,BM1}.
Here we made use of the {\tt MINIMAX} method \cite{MINIM,LANCZ}. This 
method has also been applied to derive
the analytic continuation for the heavy flavor Wilson coefficients up to 2-loop order \cite{AB}.
For another
proposal for the analytic continuation of harmonic sums to $N~\in~{\bf R}$,
for which some simple examples were presented, cf. \cite{KOT}. For other
effective parameterizations see \cite{Vogt1}.}
Since we derive the analytic continuations in analytic form, also all derivatives are 
easily determined.
In the definition of the basic functions we chose representations in the past, in which the 
functions to be Mellin-transformed may posses a branch point at $x=1$. To express the 
Mellin transforms $\Mvec[f(x)](N)$ in terms of factorial series it is required that the 
functions $f(x)$ are analytic at $x=1$, which can be obtained using well--known
mirror relations for the corresponding Nielsen integrals or harmonic polylogarithms 
transforming $x \leftrightarrow (1 - x)$. 

The paper is organized as follows. In Section~2 a brief general outline is presented on the 
way finite nested harmonic sums emerge in higher order calculations for single scale 
processes. Some basic definitions are given in Section~3. Classes of 
harmonic sums are discussed in Section~4.
In sections~5 to 9 we consider the { Mellin} transforms being associated to the
sums of weight {\sf w = 1 } to {\sf w = 5} in detail and derive the structural relations 
for the harmonic sums.~\footnote{The structural relations of {\sf w = 6} are presented
in Ref.~\cite{JB08}.} Here we will first refer to the set of 
{ Mellin} transforms chosen formerly~\cite{HS2,JB2,JBSH}. The basis found in
this way will be reordered by a linear transformation into another one which
is consistent with the representation in terms of factorial
series. Section 10 contains the conclusions. The algebraic relations to weight
{\sf w = 5} used are given in Appendix~A. Appendix~B contains the relations 
necessary to express the basic functions in the complex plane. Some useful integrals 
which emerge in this context are given in Appendix~C.

\section{The Emergence of Harmonic Sums}
%

\vspace{1mm}
\noindent
In the following we give a brief outline on the way finite nested 
harmonic sums emerge in higher order calculations for single scale processes.
These processes contain a scale, such as a momentum fraction $z$, which relates
a parton momentum $p = zP$ collinearly to the hadron momentum $P$ in the
massless limit, $0 \leq z \leq 1$. Examples for this case are space- and 
time-like splitting and
coefficient functions, cf.~\cite{ANDCO,TDYHI}. The representation also applies
to massive calculations in the limit $Q^2/m^2 \gg 1$,~~cf.~e.g.~\cite{HEAV,HEAV1}.
Likewise this variable may be viewed as the ratio of two 
Lorentz-invariants $z = t/s$, as the case for the soft and virtual corrections
to Bhabha-scattering \cite{BHABHA}, or the initial state radiation to $e^+e^-$ 
annihilation~\cite{QED}, where the respective ratio is $z=s'/s$ with $s'$ the
cms energy of the virtual gauge boson exchanged. In the latter two processes 
the electrons are massive but are dealt with in the limit $s/m^2_e \gg 1$. 
$s,s'$ and $t$ are Mandelstam-variables, i.e. Lorentz-products of linear 
combinations of 4-momenta. The factions $z$ do always obey $0 \leq z \leq 1$. 

There are two principle approaches, which can be followed in these 
calculations. Either one uses the light-cone expansion \cite{LC} for 
space-like and the cut-vertex method \cite{CUT} in the time-like case or
one follows a partonic description in the massless case, resp. the case 
$Q^2/m^2 \gg 1$, with $Q^2$ the relevant bosonic virtuality exchanged.
The light-cone expansion and cut-vertex method lead to Mellin moments 
immediately. Here, furthermore, the corresponding current crossing relations
have to be observed, which single out even or odd integer moments, cf.~\cite{BLKO}, 
depending on the physical process. In the partonic approach one may form these moments 
accordingly, transforming to Mellin space since the variables $t/s$, resp. 
$s'/s~\in~[0,1]$ in the limit mentioned above. Further on we will consider 
the representations in Mellin space only. 

The no-scale quantities can be 
obtained as a special case out of single scale quantities setting $N=const.$ or
$N \rightarrow \infty$, respectively. If in the former case, the 
representation appears in terms of nested harmonic sums, partly decorated with 
multiple $\zeta$-values, the latter are just multiple $\zeta$-values. Here
we always refer to the alternating harmonic sums and multiple $\zeta$-values,
as the present level of physics calculations for single scale processes contains these 
quantities only.
These structures are obtained by iterated integrals over the alphabet 
${1/x,~1/(1-x),~1/(1+x)}$ and their Mellin transforms.
Going to even higher orders, or including a scale more, this alphabet will be 
extended, cf.~\cite{ANDRE}, which leads to similar structures. The 
r\^ole of the harmonic polylogarithms \cite{VR} will be taken by the 
polylogarithms over the {\sf extended} alphabet. The multiple $\zeta$-values
are then generalized by the values of the new polylogarithms at $x=1$, resp. 
the symbolic polynomial in case the divergent symbol $\sigma_0$ appears.
The nested harmonic sums are generalized into linear combinations of Mellin 
transforms of the new polylogarithms, which one may call {\sf generalized 
harmonic sums}. 

Let us now describe how these structures emerge. We consider a renormalizable
Quantum Field Theory in $D = n + \varepsilon$ space--time dimensions. In case
of the Standard Model and its parts, Quantum Electrodynamics and Quantum 
Chromodynamics, the dimension is $n=4$. The propagators and vertices of the 
Feynman diagrams are combined with the help of Feynman parameters, 
\begin{eqnarray}
\frac{1}{A^\alpha \cdot B^\beta} = \frac{\Gamma(\alpha + \beta)}{\Gamma(\alpha)
\Gamma(\beta)}
\int_0^1 
\frac{dx~~x^{\alpha-1}~(1-x)^{\beta-1}}{[x A + (1-x)B]^{\alpha+\beta}}~.
\end{eqnarray}
This allows to re-express the otherwise difficult angular integrals. In this 
way the loop $D$--momenta are synchronized. The Wick-rotation is performed, 
which leads to Euclidean integrals for the loop momenta in $D=n+\varepsilon$
dimensions. The numerator integrals may contain operator insertions like
\begin{eqnarray}
\label{opins}
(\Delta.k_i)^m~,
\end{eqnarray}
and several finite sums over these terms in $m$, which are bounded by the 
Mellin variable $N$. The operator insertion (\ref{opins}) in momentum space 
stem from local 
composite operators emerging in operator product- or light cone expansions.  
Here $k_i$ denotes a loop momentum, $m$ an 
integer and $\Delta$ a light-like 
$D$-vector with $\Delta.\Delta = 0$. The dot denotes the 
Minkowski-product, which turns into the scalar product through Wick-rotation.
The trivial angular integrals and radial integrals over the loop 
momenta can now be performed, which yield a factor 
\begin{eqnarray}
\left.
\frac{\Gamma(n_1 + r_1 \varepsilon) \ldots \Gamma(n_k + r_k \varepsilon)}
     {\Gamma(m_1 + q_1 \varepsilon) \ldots \Gamma(m_l + q_l 
\varepsilon)}\right|_{n_i,m_j~\in~{\bf Z}, r_i,q_j~\in~{\bf Q}}~,
\end{eqnarray}
leaving the non-trivial Feynman parameter integrals. In carrying out theses integrals 
scalar and tensor integrals have to be performed.

The Feynman parameter integrals contain the Mellin variable in the 
numerator only. The denominator is given by the Kirchhoff-polynomial of the 
respective graph 
consisting out of $k$ monomials raised to a real non-integer power. A possible 
next step 
consists in expressing the Feynman-parameter integrals in terms of Mellin-Barnes 
integrals \cite{MELBAR} for each of these monomials, 
\begin{eqnarray}
\frac{1}{(A+B)^q} = \frac{1}{2 \pi i} 
\int_{\gamma-i\infty}^{\gamma+i\infty}~d\sigma~A^\sigma~B^{-q-\sigma} 
\frac{\Gamma(-\sigma) \Gamma(q+\sigma)}{\Gamma(q)}~.
\end{eqnarray}
The use of Mellin-Barnes integrals usually leads to rather large set of sums
if compared to the sums needed to express the physics results. We refer to 
this method, since it uniquely allows to create the structures 
emerging.~\footnote{A much more economic way consists in representing 
the different contributions to the Feynman parameter integrals in terms of 
generalized hypergeometric and related functions {\it directly}, 
cf.~[6d].}
All the Feynman-parameter integrals can now be performed and lead to a further 
factor of the kind
\begin{eqnarray}
\left.
\frac{
 \Gamma(a_1 N + b_1(\sigma_a) + \bar{r}_1 \varepsilon) 
 \ldots
 \Gamma(a_k N + b_k(\sigma_a) + \bar{r}_k \varepsilon)} 
{\Gamma(c_1 N + d_1(\sigma_a) + \bar{q}_1 \varepsilon) 
 \ldots
 \Gamma(c_l N + d_l(\sigma_a) + \bar{q}_l \varepsilon)}\right|_{a_i ... 
d_i~\in~{\bf Z}, 
\bar{r}_i,\bar{q}_j~\in~{\bf Q}}~.
\end{eqnarray}
Here $b_i,d_i$ are linear functions of the Mellin--Barnes variables with 
integer coefficients.

Now the $k$ Mellin--Barnes integrals have to be performed. This can be done
applying the residue theorem, after the according contours have to be 
properly fixed. In this way one obtains $k$ or more infinite sums over the
above rational function of $\Gamma$-functions, with according replacement of 
the sum-index in place of the variables $\sigma_a$. 

For these multiple sum expressions one may seek representations which are 
generalized hypergeometric series \cite{HYP,HEAV1} and generalizations thereof.
Going to ever higher orders not all of these functions may be known by now
but have to be introduced newly. They are naturally generated by the above 
integrals.

At this stage one may wish to expand in $\varepsilon$. If quantities without 
infrared singularities are considered, the maximal negative power in 
$\varepsilon$ is given by the loop-order of the diagram. Infrared 
singularities may enhance this order. The expansion has to be carried out 
up to the respective order required in the calculation. As only 
$\Gamma$--functions are expanded, which partly contain $a_l k_l$, with 
$a_l,k_l~\in~{\bf N}$, single harmonic sums of the type $S_b(a_l k_l)$ 
emerge 
in products. The various infinite sums with index $k_l$ nest these products,
which finally leads to {\sf nested harmonic sums} and {\sf multiple $\zeta$-values} or 
their respective generalizations in higher orders, as described 
above.~\footnote{First non-harmonic sums were found in individual diagrams in 
\cite{COEF3}. They canceled, however, in the final result in the 
$\overline{\rm MS}$-scheme within each 
color factor.} The according sums can be uniquely found investigating the 
corresponding recurrences in $N$ over $\Pi \Sigma$--fields, as possible with 
the code {\tt Sigma},~\cite{SIGMA}.

The harmonic sums, which occur in the calculations of single scale quantities
up to a given weight can be represented over a basis of functions, which is 
independent of the respective quantity being considered. In the following
we identify these basis elements up to weight {\sf w = 5}.
\section{Basic Definitions}
%

\vspace{1mm}\noindent
We will consider one--parameter real--valued functions $x~\in~]0,1[$.  
Partly this set will be extended to special distributions $f(x)~\in~{\cal D}'(]0,1[)$
\cite{DIST} occurring in field theoretic  calculations. Their { Mellin} 
transform is defined by Eq.~(\ref{eqMELL}),
where $N$ is a positive, suitably large integer. According to the emergence 
of 
the respective { Mellin} transform in quantum field theoretic calculations, 
such as the local light cone expansion \cite{LC}, $N$ is either an even or odd integer. 
Factors of $(-1)^N$, 
which emerge in the following, have therefore a definite 
meaning and are either equal to $+1$ or $-1$. Both branches may be 
continued 
analytically in $N$ from $\Nvec$ to $\Cvec$ according to Carlson's theorem 
\cite{CARLS}. The distributions $f(x)$ considered in the following
are differentiable functions in the class ${\cal C}^{\infty}(]0,1[)$ or
$\delta$--distributions $\delta^{(k)}(1~-~x),~k \geq 0$ in ${\cal D'}[0,1]$~\cite{DIST}.
Furthermore, we consider functions with a $+$-prescription,  
\begin{eqnarray}
\int_0^1 dx~ x^{N}  \left[f(x)\right]_+ = \int_0^1 dx~ \left(x^{N}
-1 \right) f(x)~,  
\end{eqnarray}
with $f~\in~{\cal C}^{\infty}(]0,1[)$. The { Mellin} transform of the
latter functions are harmonic sums in case the functions $f(x)$ are
harmonic polylogarithms~\cite{VR} weighted by a factor $1/(1 \pm x)$, which we will 
consider throughout the present paper.
The { Mellin} transform of the $\delta^{(k)}(1-x)$--distributions yield
\begin{eqnarray}
\int_0^1 dx~ x^{N}  \delta^{(k)}(1-x) = (-1)^k \prod_{l=0}^k (N-l)~.
\end{eqnarray}
In all cases below which are not of the form $f(x)=\ln^m(x) \hat{f}(x)$, or a linear 
combination of these terms, the function 
can be expanded into a { Taylor} series for $x~\in~[0,1[$. 
\begin{eqnarray}
f(x) = \sum_{k=0}^\infty a_k x^k,~~~~~a_k = \frac{f^{(k)}(0)}{k!}~,~
\forall a_k~\in~\Rvec~.
\end{eqnarray}
The { Mellin} transform of $f(x)$ is
\begin{eqnarray}
\int_0^1 dx~ x^{N} f(x) = \sum_{k=0}^{\infty} \frac{a_k}{N+1+k}~.   
\end{eqnarray}
All the above { Mellin} transforms are meromorphic functions in $N$.
For the case $f(x) = \ln^m(x) \hat{f}(x)$ 
\begin{eqnarray}
\int_0^1 dx~ x^{N} \ln^m(x) \hat{f}(x) = \frac{\partial^m}{\partial N^m} 
\int_0^1 dx~ x^{N} \hat{f}(x)
\end{eqnarray}
holds accordingly.

The class of functions $\hat{f}(x)$ is formed out of polynomials of harmonic polylogarithms
\cite{VR}. For physical applications in the massless case it will turn out 
that all contributing harmonic sums up to weight {\sf w = 5} can be 
represented by polynomials
out of Mellin transforms of the type
\begin{eqnarray}
\Mvec\left[\left(\frac{\hat{f}(x)}{x-1}\right)_+\right](N),~~~~~~~{\rm or}~~~~~~~~~
\Mvec\left[\frac{\hat{f}(x)}{x+1}\right](N)~,
\end{eqnarray}
where $f(x)$ is a polynomial of { Nielsen} integrals. In the analytic continuation
one may consider Mellin transforms at rational arguments, like $N/2$ and 
study the mapping
\begin{eqnarray}
\label{eqx2x}
\Mvec\left[\frac{\hat{f}(x^2)}{1-x^2}\right](N) \rightarrow 
\Mvec\left[\frac{\hat{f}(x)}{1-x}\right]\left(\frac{N-1}{2}\right)~,  
\end{eqnarray}
for $\hat{f}(1) = 0$.
The factor decomposition of the denominators yields
\begin{eqnarray}
\frac{1}{1-x^2} = \frac{1}{\Phi_1(x) \Phi_2(x)} 
= \frac{1}{2} \left[\frac{1}{1+x} 
                  + \frac{1}{1-x}\right]~,
\end{eqnarray}
where $\Phi_{1,2}(x)$ denote the first two cyclotomic 
polynomials~\cite{CYCL}~\footnote{
In massive calculations higher cyclotomic polynomials may occur as well.
Examples are the fractional values of the $\beta$-function (\ref{eqBET0}), as
$\beta((x+1)/2) = 2 \Mvec[1/(1+z^2)](x),~~~\beta((x+1)/3) = \beta(x+1)- 
\Mvec[(z-2)/(z^2-z+1)](x)$  for the 4th and 6th cyclotomic 
polynomial, cf.~\cite{NIELS2}. See also \cite{WEINZ}.}.
The representation of  a harmonic polylogarithm $f(x^2)$ in terms of other
harmonic polylogarithms $f_i(x)$ may be worked out for non-negative 
indices, \cite{VR}. We will give 
the explicit relation for the case of the { Nielsen} integrals 
$S_{n,p}(x)$ \cite{NIELS1}
\begin{eqnarray}
S_{n,p}(x) = \frac{(-1)^{n+p-1}}{(n-1)! p!}
\int_0^1 \frac{dz}{z} \ln^{n-1}(z) \ln^p(1-xz)~.
\end{eqnarray}
The following relation holds
\begin{eqnarray}
S_{n,p}(x^2) =  2^{n} \frac{(-1)^{n+p-1}}{(n-1)!p!} 
\sum_{l=0}^p \binom{p}{l} \int_0^x 
\frac{dz}{z} \ln^{n-1}\left(\frac{z}{x}\right) \ln^{p-l}(1-z) \ln^l(1+z)~. 
\end{eqnarray}
For $n=1$ one obtains
\begin{eqnarray}
\frac{1}{2} S_{1,p}(x^2) =  S_{1,p}(x) + S_{1,p}(-x) + \frac{(-1)^{p}}{p!} 
\sum_{l=1}^{p-1} \binom{p}{l} \int_0^1 
\frac{dz}{z}  \ln^{p-l}(1-zx) \ln^l(1+zx)~. 
\end{eqnarray}
Even in this simple case for Nielsen integrals, which are not 
classical polylogarithms, the mapping of the argument 
$x \rightarrow x^2$ does not give a closed relation in the same class 
of functions anymore. However, relations of this type may be used to express
certain Mellin transforms, cf. \cite{HS2}, for examples  up to {\sf w = 4}.

\section{Classes of Harmonic Sums}
%

\vspace{1mm}
\noindent
The number of all possible alternating and non--alternating harmonic sums
of weight {\sf w}, $N({\sf w})$, is
\begin{equation}
\label{eqNW1}
N({\sf w}) = 2 \cdot 3^{{\sf w} - 1}~. 
\end{equation}
The number of basic sums in the case of all permutations can be calculated 
using the 1st Witt formula \cite{WITT} for an alphabet of length {\sf l = 3}
for {\sf w $> 1$},~\cite{JBSH},
\begin{eqnarray}
N^{\rm basic}({\sf 1}) &=& 2,\nonumber\\
\label{eqbas1}
N^{\rm basic}({\sf w}) &=& \frac{1}{{\sf w}} \sum_{d|{\sf w}} 
\mu\left(\frac{{\sf w}}{d}\right) 3^d,~~~~{\sf w}~\geq~2~.
\end{eqnarray} 
The alternating and non--alternating harmonic sums 
\cite{HS3,HS2}
are generated by Mellin 
transforms of nested integrals based on the three {\sf w = 1} functions
\begin{equation}
\frac{1}{1+x}~, \hspace{1cm} \frac{1}{x}~, \hspace{1cm} \frac{1}{1-x}~.
\end{equation}
Therefore, the alphabet out of which the index of the respective harmonic 
sums is formed as a word is of length {\sf l = 3}. The harmonic sums
form a sub--set w.r.t all possible index--sets over the alphabet $\{-1, 0,
1 \}$, since the letter $0$ cannot appear first. The counting relation is 
given by Eq.~(\ref{eqNW1}).

Investigating the structure of the single--scale quantities in QCD and 
QED, cf.~ \cite{HEAV,HEAV1,COHS,JB04,BM1,HL}
to 2-- and 3--loop order in more detail, it 
turns out that harmonic sums with index $i = \{-1\}$ do not contribute. 
One finds that the number of all harmonic sums of weight {\sf 
w}, which do not 
contain any index $i = -1$, is obtained by expanding the following 
generating function~\cite{SEQ1}
\begin{equation}
\frac{1-x}{1 - 2 x - x^2} = \sum_{{\sf w} = 0}^{\infty} N_{\neg 
(-1)}({\sf w}) x^{\sf w}~,
\end{equation}
with
\begin{eqnarray}
N_{\neg \{-1\}}({\sf w}) &=& \frac{1}{2} \left[ 
\left(1-\sqrt{2}\right)^{\sf w} 
+ \left(1+\sqrt{2}\right)^{\sf w} \right]
= \sum_{k=0}^{[w/2]}~\binom{w}{2k}~2^k~.
\end{eqnarray}
$N_{\neg \{-1\}}({\sf w})$ obeys the recursion relation
\begin{eqnarray}
N_{\neg \{-1\}}({\sf w}) = 2 \cdot N_{\neg \{-1\}}({\sf w-1}) + N_{\neg 
\{-1\}}({\sf w-2})
\end{eqnarray}
with the starting values $N_{\neg \{-1\}}({\sf 1}) = 1,~N_{\neg \{-1\}}({\sf 2}) = 3$~.
Another interesting sub--class of the harmonic sums is that
of the non--alternating harmonic sums. 
These sums are formed over an alphabet of length {\sf l~=~2}~\cite{HOFF}. Similarly  to 
(\ref{eqbas1}) one obtains 
\begin{eqnarray}
\label{eqNW0}
N_{i~>~0}({\sf w}) &=&  2^{{\sf w} - 1}~, \\
N_{i~>~0}^{\rm basic}({\sf 1}) &=& 1~,\nonumber\\
\label{eqbas0}
N_{i~>~0}^{\rm basic}({\sf w}) &=& \frac{1}{{\sf w}} \sum_{d|{\sf w}} 
\mu\left(\frac{{\sf w}}{d}\right) 2^d,~~~~{\sf w}~\geq~2~~.
\end{eqnarray} 

In Table~1 the number of sums is given in dependence of the 
weight {\sf w}. Here the number of a-basic sums is the number of sums 
obtained using the algebraic relations  \cite{JBSH}.~\footnote{Later on we
will derive a basis representation for the harmonic sums including the structural 
relations. Since in this representation we make a special choice,
we list the corresponding algebraic relations w.r.t. to this basis for the
dependent sums in Appendix~A.}
As shown in Table~1 the number of a-basic sums for all harmonic sums,
$N_{\rm basic}({\sf w})$, is larger or equal then the number of all
harmonic sum, which do not contain $\{-1\}$ as an index. 
The corresponding number of basic sums is given by~:
\begin{eqnarray}
N^{\rm basic}_{\neg \{-1\}}({\sf w})~=~\frac{2}{{\sf w}} \sum_{d|{\sf w}}
\mu\left(\frac{{\sf w}}{d}\right) N^{\rm basic}_{\neg \{-1\}}(d)~.
\end{eqnarray}
Values of $N^{\rm basic}_{\neg \{-1\}}({\sf w})$
are partly connected to the number of a non--rooted tree studied in
\cite{TREE} for {\sf w}$~=~2 \ldots 7$, but obey a different counting
rule otherwise.

We mention that relation (\ref{eqbas1}) applies also to other objects, as 
the harmonic polylogarithms, where the respective numbers up to {\sf w = 8} 
were determined empirically in \cite{VR}. Here we may apply it even for {\sf w = 1}, 
unlike the case for harmonic sums.

\begin{center}
\renewcommand{\arraystretch}{1.3}
\begin{tabular}[h]{||r||r|r||r|r||r|r||}
\hline \hline %
\multicolumn{1}{||c||}{ } &
\multicolumn{6}{c||}{\sf Number of } \\
\cline{2-7}
\multicolumn{1}{||c||}{\sf Weight}&
\multicolumn{1}{c|}{\sf Sums}&
\multicolumn{1}{c||}{\sf a-basic sums }&
\multicolumn{1}{c|}{\sf Sums $\neg \{-1\}$}&
\multicolumn{1}{c||}{\sf a-basic sums} &
\multicolumn{1}{c|}{\sf Sums $i > 0$}  &
\multicolumn{1}{c||}{\sf a-basic sums} \\
\hline\hline
  1 &     2 &        2 &    1 &   1  &   1 &  1 \\
  2 &     6 &        3 &    3 &   2  &   2 &  1 \\
  3 &    18 &        8 &    7 &   4  &   4 &  2 \\
  4 &    54 &       18 &   17 &   7  &   8 &  3 \\
  5 &   162 &       48 &   41 &  16  &  16 &  6 \\
  6 &   486 &      116 &   99 &  30  &  32 &  9 \\
  7 &  1458 &      312 &  239 &  68  &  64 & 18 \\
  8 &  4374 &      810 &  577 & 140  & 128 & 30 \\
  9 & 13122 &     2184 & 1393 & 308  & 256 & 56 \\
 10 & 39366 &     5880 & 3363 & 664  & 512 & 99 \\ 
\hline \hline
\end{tabular}
\renewcommand{\arraystretch}{1.0}
\end{center}

\vspace{2mm} \noindent
{\sf Table~1:~Number of harmonic sums, number of sums, which do not
contain the index $\{-1\}$, number of  sums with positive indices and 
the respective numbers of basic sums by which all sums can be expressed 
using the algebraic relations (a-basic sums)
\cite{JBSH} in dependence of their weight.}
 
\section{Sums of weight one}
%

\vspace{1mm}
\noindent
The harmonic sums of weight {\sf w=1} are generated by the  Mellin
transform of the
first two cyclotomic polynomials~\cite{CYCL}~\footnote{These polynomials are
the real--valued factors of $x^N-1$ with $N \geq 1, N \in {\bf \rm N}$.} 
(after regularization of the 
integral)  
\begin{eqnarray}
\frac{1}{1-x} \hspace*{1.5cm} {\rm and}
\hspace{1.5cm} \frac{1}{1+x}
\end{eqnarray}
by
\begin{eqnarray}
\label{LEG}
\Mvec\left[\left(\frac{1}{1-x}\right)_+\right](N) &=& -S_1(N) =
-\psi(N+1)-\gamma_E \\
\Mvec\left[\frac{1}{1+x}\right](N) &=& (-1)^{N}\left[S_{-1}(N) + \ln(2)
\right] = \beta(N+1)~.
\end{eqnarray}
Here $\gamma_E$ denotes the { Euler--Mascheroni} constant.
The representation of the integral in Eq.~(\ref{LEG}) in terms of the
$\psi$--function was first given by { Legendre}~\cite{LEGAN}.
The { Mellin} transform for an integrable function $x^N f(x)$ obeys the identity
\begin{eqnarray}
\Mvec\left[f\left(x^a\right)\right](N) &=& \frac{1}{a}
\Mvec[f(x)]\left(\frac{N+1-a}{a}\right),~~a~\in~\Rvec,~~a~>~0~.
\end{eqnarray}
This relation modifies in case of $f(x)$ being a $+$-distribution.
For $a=2$ one obtains
\begin{eqnarray}
\label{eqDEC1}
\frac{1}{1-x^2} &=&  \frac{1}{2}\left[\frac{1}{1-x} + \frac{1}{1+x}
\right]\\
\Mvec\left[\left(\frac{1}{1-x}\right)_+\right]\left(\frac{N-1}{2}\right)
&=&
\Mvec\left[\left(\frac{1}{1-x}\right)_+\right](N) +
\Mvec\left[\frac{1}{1+x}\right](N) +  \ln(2)~.
\end{eqnarray}

The latter relation can even be analytically continued in closed form  
since, see \cite{HS2},
\begin{eqnarray}
-\psi\left(\frac{N}{2}\right) -\gamma_E
&=& -\psi(N) -\gamma_E +\beta(N) +\ln(2)~,
\end{eqnarray}
with
\begin{eqnarray}
\label{eqBET0}
\beta(N) = \frac{1}{2} \left[\psi\left(\frac{N+1}{2}\right) - \psi\left(
\frac{N}{2}\right)\right]~,
\end{eqnarray}
cf.~\cite{STIRLING,NIELS2}. The duplication relation for the $\psi$--function 
yields
\begin{eqnarray}
\psi(N) &=& \frac{1}{2}\left[\psi\left(\frac{N}{2}\right) +
\psi\left(\frac{N+1}{2}\right) \right] + \ln(2)~.  
\end{eqnarray}
Here fractional arguments 
in the $\psi$--function emerge quite naturally. For $k > 1$ the harmonic sums
$S_{-k}(N)$ and $S_k(N)$ obey, respectively~: 
\begin{eqnarray}
\label{sing1}
S_{-k}(N) &=& \frac{(-1)^{(k-1)}}{(k-1)!} (-1)^N \beta^{(k-1)}(N+1) 
                - \left[1 - \frac{1}{2^{k-1}}\zeta(k)\right]\\
\label{sing2}
S_k(N)    &=& \frac{(-1)^{(k-1)}}{(k-1)!} \psi^{(k-1)}(N+1) + \zeta(k)~,
\end{eqnarray}
with $\zeta_k$ the Riemann $\zeta$--function.
All these sums are expressed by the $\psi$--function and its derivatives.
We therefore conclude that due to the above rational-argument and differential relations 
there is only one basic single harmonic sum, $S_1(N)$.

One may wonder whether differentiation relations can also be of importance for
multiple zeta values. This is indeed the case~\cite{CARTIER}. The multiple zeta values 
can be generalized to multiple Hurwitz zeta values
\cite{HURWITZ} defined  by
\begin{eqnarray}
\label{HURW}
\zeta^a_{c,\vec{d}} = \sum_{k_1 = 1}^\infty \frac{({\rm sign}(c))^{k_1}}{(a + 
k_1)^{|c|}} S^{H,a}_{\vec{d}}(k_1)~,
\end{eqnarray}
iteratively with
\begin{eqnarray}
\label{HURW1}
S^{H,a}_{\vec{d}}(k_1) = \sum_{k_2=1}^{k_1} \frac{({\rm 
sign}(d_1))^{k_2}}{(a+k_2)^{|d_1|}} S^{H,a}_{d_2 ... d_m}(k_2)~.
\end{eqnarray}
Here $a~\in~{\bf C}$ so that the above sums are defined.
One may differentiate for $a$ and perform then  the limit $a \rightarrow 0$.
Multiple Hurwitz zeta values obey stuffle relations, using the diction of 
Ref.~\cite{LISO}. Furthermore, one may form shuffle, duplication and 
more general algebraic relations among them, similar to the case 
studied for the multiple zeta values in Ref.~\cite{BBV}. Differentiating
these relations builds classes of these relations which may be  
somewhat easier  obtained as working out the relations directly for 
the multiple zeta values. It is not excluded that at very high weights 
even more relations between multiple zeta values can be found than obtained in
\cite{BBV}, although for lower weights, i.e. {\sf w $\leq 12$}  this is not 
expected. 

The differential quotient of a harmonic sum is always of the form
\begin{eqnarray}
\label{diff1}
\frac{d}{dN} S_{\vec{a_1}}(N) = S_{\vec{a_2}}(N)- 
\zeta_{\vec{a_2}},~~~~\lim_{N 
\rightarrow \infty} S_{\vec{a_2}}(N) =  \zeta_{\vec{a_2}}~.
\end{eqnarray}
This is due to the fact that (\ref{diff1}) can be represented by a Mellin 
transform which contains $x^N$ as a factor of the integrand being continuous 
in $x~\in~]0,1]$. The Mellin transform vanishes for $N 
\rightarrow \infty$. 

As an example we consider single harmonic sums. Differentiation yields
\begin{eqnarray}
\frac{d}{dN} S_k(N)    &=& - k \left[S_{k+1}(N) - \zeta(k+1)\right]
\\
\frac{d}{dN}\left[(-1)^N S_k(N)\right]    &=& - (-1)^N k \left[S_{-(k+1)}(N) + 
\left(1-\frac{1}{2^{k-1}}\right) 
\zeta(k+1)\right]
\end{eqnarray}
Differentiating the Hurwitz zeta value $\zeta^a(k_1)$ and taking the 
limit $a \rightarrow 0$ yields
\begin{eqnarray}
\lim_{a \rightarrow 0} \frac{d}{da} \zeta^a_c &=& -c~\zeta_{+(c+1)},~~~~~c>0 \\ 
\lim_{a \rightarrow 0} \frac{d}{da} \zeta^a_c &=& \hspace{3mm} 
c~\zeta_{-(c-1)},~~~~~c<0~.  
\end{eqnarray}

\section{Sums of weight two}
%

\vspace{1mm}
\noindent
Among the six harmonic sums four decompose into polynomials of single sums,
which may be expressed through the $\psi^{(k)}(z)$ functions as shown in
the previous section. Two sums, $S_{-1,1}(N)$ and $S_{1,-1}(N)$ remain. Due
to the simplest shuffle relation by { Euler}~\cite{EUL}
\begin{eqnarray}
\label{Eqeul}
S_{a,b}(N) + S_{b,a}(N) =  S_a(N) S_b(N) + S_{a \wedge b}(N) 
\end{eqnarray}
with
\begin{eqnarray}
a \wedge b = {\rm sign}(a) {\rm sign}(b) \left[|a| + |b|\right]
\end{eqnarray}
these sums are connected, which implies the following relation for the {
Mellin} transforms
\begin{eqnarray}
\Mvec\left[\frac{\ln(1-x)}{1+x}\right](N) &=&
-\Mvec\left[\frac{\ln(1+x)}{1+x}\right](N) \nonumber\\
& &+ (-1)^{N+1}\left[S_1(N) S_{-1}(N) + S_{-2}(N) \right.\nonumber\\
& &+ \left. \left[S_1(N)-S_{-1}(N)\right] \ln(2) 
- \ln^2(2) + \frac{1}{2} \zeta_2 \right]~,
\end{eqnarray}
cf.~\cite{HS2}.
Similar to (\ref{eqDEC1}) one may decompose
\begin{eqnarray}
\label{eqlnx2}
\frac{\ln(1-x^2)}{1-x^2} &=& \frac{1}{2} \left\{
  \frac{\ln(1-x)}{1-x}
+ \frac{\ln(1-x)}{1+x}
+ \frac{\ln(1+x)}{1-x}
+ \frac{\ln(1+x)}{1+x} \right\}~,
\end{eqnarray}
which yields
\begin{eqnarray}
\label{eqA}
\Mvec\left[\left(\frac{\ln(1-x)}{1-x}\right)_+\right]\left(\frac{N-1}{2}\right)
&=&
~~~~~\Mvec\left[\left(\frac{\ln(1-x)}{1-x}\right)_+\right](N)
+ \Mvec\left[\frac{\ln(1-x)}{1+x}\right](N)  \nonumber\\ & &
+ \Mvec\left[\frac{\ln(1+x)}{1+x}\right](N)
+ \Mvec\left[\left(\frac{\ln(1+x)}{1-x}\right)_+ \right](N)
\nonumber\\ &&
-\frac{1}{2} \zeta_2 + \ln^2(2)~. 
\end{eqnarray}
Since the l.h.s. of (\ref{eqA}) and the first and fourth term in the 
r.h.s. are
polynomials of single sums, the remainder two terms are related. However, 
(\ref{eqlnx2}) does not yield a new relation. Therefore
\begin{eqnarray}
\label{eqT1}
F_{1}(N) : = \Mvec\left[\frac{\ln(1+x)}{1+x}\right](N)
\end{eqnarray}
is the first  non-trivial { Mellin} transform beyond the single harmonic sum
$S_1(N)$. Although $F_1(N)$ 
expresses a harmonic sum containing $\{-1\}$ as index, which belongs to a class
being absent in the final results in physics applications, it may be of use
to derive compact expressions in other cases later. 
 
As a historic remark, we mention that { Nielsen}~\cite{NIELS2}
considered these functions as well using the notation $\xi(N), \eta(N), 
\xi_1(N)$ and $\xi_2(N)$,
\begin{eqnarray}\
\xi(N) &=& \Mvec\left[\left(\frac{\ln(1-x)}{x-1}\right)_+\right](N-1)
= \frac{1}{2} \left[\psi'(N)-\zeta_2
-\left(\psi(N)+\gamma_E\right)^2\right] 
\nonumber\\ &=&
-S_{1,1}(N-1)
\\
\eta(N) &=& \Mvec\left[\frac{\ln(1+x)-\ln(2)}{x-1}\right](N-1)
\\
\xi_1(N) &=& \Mvec\left[\frac{\ln(1+x)}{x+1}\right](N-1)
\\
-\xi_2(N) &=& \Mvec\left[\frac{\ln(1-x)}{x+1}\right](N-1)~.
\end{eqnarray}
They obey the relations
\begin{eqnarray}
\left[\psi(z)+\gamma_E\right]\left[\psi(1-z)+\gamma_E\right]
&=& 2 \zeta_2 -\xi(z) - \xi(1-z)
\\
\label{eqbeps}
\beta(z) \left[ \psi(1-z) + \gamma_E \right] &=& 
- \beta(1-z) \ln(2) - \xi_1(z) - \xi_2(1-z)
\\ 
\label{eqbeps1}
\beta(z) \left[ \psi(z) + \gamma_E \right] &=& 
 \beta'(z) + \beta(z) \ln(2) + \frac{1}{2} 
\xi\left(\frac{z}{2}\right) - \frac{1}{2} \xi \left(\frac{z+1}{2}\right)
\\
\beta(z)+\beta(1-z) &=& \frac{\pi}{\sin(\pi z)}
\\
\beta(z) \beta(1-z) &=& \eta(z) + \eta(1-z)   
\\
\beta^2(z) &=& \psi'(z) - 2 \eta(z) 
~,
\end{eqnarray}
for $z~\in~]0,1[$. Most of them were given in Ref.~\cite{NIELS2}. They 
result from algebraic relations between
harmonic sums and corresponding integral representations. Let us illustrate 
this for (\ref{eqbeps}). One may represent
\begin{eqnarray}
\beta(N+1) \left[\psi(N+1)+\gamma_E\right] = (-1)^N S_1(N) 
\left[S_{-1}(N)+\ln(2)\right],
\end{eqnarray}
cf.~\cite{HS2}. Using Euler's relation (\ref{Eqeul}) one obtains
\begin{eqnarray}
\beta(N+1) \left[\psi(N+1)+\gamma_E\right] = 
(-1)^N\left[S_{1,-1}(N) + S_{-1,1}(N) - S_{-2}(N) + \ln(2) S_1(N)\right]~,
\end{eqnarray}
which can be expressed referring to the functions $\beta(N), \beta'(N), 
\xi_1(N)$ and $\xi_2(N)$. Relation (\ref{eqbeps1}) is a consequence of the 
properties of the logarithm.~\footnote{
We correct misprints in Ref.~\cite{NIELS2},~\S 75, third equation. The minus 
sign in front of $\beta'(x)$ has to be removed. The equation below Eq. (8) has to 
be replaced by (\ref{eqbeps}).}
\section{Sums of weight three}
%

\vspace{-1mm}
\noindent
There are eighteen harmonic sums of weight {\sf w = 3},~\cite{HS2}. Twelve of 
these 
sums 
are related to the remaining six harmonic sums by shuffle
relations~\cite{JBSH}. The latter are represented by the { Mellin}
transforms
\renewcommand{\arraystretch}{1.3}
\begin{equation}
\label{eq5a}
\begin{array}{lll}
\Mvec\left[\ds \left(\frac{\ln^2(1+x)}{1-x}\right)_+\right](N) &
\hspace{2.0cm}
& 
\Mvec\left[\ds \frac{\ln^2(1+x)}{1+x}\right](N) \\
& & \\
\Mvec\left[\left(\ds \frac{\Li_2(-x)}{1-x}\right)_+\right](N) &
\hspace{2.0cm}
&
\Mvec\left[\left(\ds \frac{ \Li_2(x)}{1-x}\right)_+\right](N) \\
& & \\
\Mvec\left[\ds \frac{\Li_2(-x)}{1+x}\right](N) & \hspace{2.0cm} &
\Mvec\left[\ds \frac{\Li_2(x)}{ 1+x}\right](N)~. \\
\end{array}
\end{equation}
\renewcommand{\arraystretch}{1.0}

\noindent
The first two functions in (\ref{eq5a}) correspond to the harmonic sums
of the type $S_{-1,1,-1}(N)$ and $S_{1,1,-1}(N)$, which are not connected 
algebraically. Using the general relation
\begin{eqnarray}
\label{eqDER}
\Mvec\left[\ln^l(x) f(x)\right](N) = \frac{d^l}{d N^l} \Mvec\left[
f(x)\right](N)~,
\end{eqnarray}
the { Mellin} transform $\Mvec\left[\ln(x) \ln(1+x)/(1+x)\right](N)$
can be calculated from $F_1(N)$. We exploit { Euler's} relation 
for the sums $S_{1,-2}(N)$ and $S_{-2,1}(N)$, \cite{HS2}, to express
$\Mvec\left[\Li_2(-x)/(1+x)\right](N)$ in terms 
of $\Mvec\left[\Li_2(x)/(1+x)\right](N)$,
\begin{eqnarray}
\label{R2P}
\Mvec\left[\frac{\Li_2(-x)}{1+x}\right](N) &=& 
(-1)^{N+1} \Biggl\{\ln(2) 
\left[S_2(N)-S_{-2}(N)\right]+\frac{1}{2} \zeta_2 S_{-1}(N) + 
\frac{3}{4} \zeta_3 
\nonumber\\ && 
+ S_{-1}(N) S_2(N) + S_{-3}(N)\Biggr\}
 + \Mvec\left[\frac{\Li_2(x) + \ln(x) \ln(1-x)}{1+x}\right](N) 
\nonumber\\
\end{eqnarray}
Furthermore, the identity
\begin{eqnarray}
\label{LIREL}
\frac{1}{2^{n-2}}\frac{\Li_n(x^2)}{1-x^2} &=&
\frac{\Li_n(x)}{1-x}
+ \frac{\Li_n(x)}{1+x} + \frac{\Li_n(-x)}{1-x} + \frac{\Li_n(-x)}{1+x}
\end{eqnarray}
holds and one obtains 
\begin{eqnarray}
\label{eq55}
\Mvec\left[\left(\frac{\Li_2(-x)}{x-1}\right)_+\right]\left( N \right)
&=&  \frac{1}{2}
\Mvec\left[\left(\frac{\Li_2(x)}{x-1}\right)_+\right]\left(\frac{N-1}{2}\right)
+\Mvec\left[\frac{\Li_2(-x)}{x+1}\right](N) 
\nonumber\\ & &
-\Mvec\left[\left(\frac{\Li_2(x)}{x-1}\right)_+\right]\left( N \right)
+\Mvec\left[\frac{\Li_2(x)}{x+1}\right]\left( N \right)
\nonumber\\ & &
-\frac{3}{8} \zeta_3 + \frac{1}{2} \zeta_2 \ln(2)~.
\nonumber\\ 
\end{eqnarray}
Eq.~(\ref{eq55}) can be used to express 
$\Mvec\left[\left(\Li_2(-x)/(1-x)\right)_+\right](N)$ in terms of the 
remaining { Mellin} transforms.

Four basic functions emerge at weight {\sf w = 3}
\begin{eqnarray}
F_2(N) &:=& 
\Mvec\left[\frac{\ln^2(1+x)-\ln^2(2)}{1-x}\right](N)\\
F_3(N) &:=& \Mvec\left[\frac{\ln^2(1+x)}{1+x}\right](N)\\
F_4(N) &:=& \Mvec\left[\left(\frac{\Li_2(x)}{1-x}\right)_+\right](N)\\ 
F_5(N) &:=& \Mvec\left[\frac{\Li_2(x)}{1+x}\right](N)~.
\end{eqnarray}
Here we have chosen a form for $F_2(N)$ which yields a particular
simple asymptotic representation as $N~\in~{\bf C},~~|N| 
\rightarrow \infty$.
The basic sums at weight {\sf w = 3} which do not contain the index $\{-1\}$
are thus given by
\begin{eqnarray}
S_{-2,1}(N) &=&
(-1)^{N+1} F_5(N) + \zeta_2 S_{-1}(N) - \frac{5}{8} \zeta_3 + \zeta_2 \ln(2)
\\
S_{2,1}(N) &=&
 F_4(N) + \zeta_2 S_{1}(N)~. 
\end{eqnarray}
\section{Sums of weight four}
%

\vspace{1mm}
\noindent
Out of the 54 harmonic sums which emerge at weight {\sf w = 4},~~~38 harmonic sums 
can
be expressed by 16 sums through shuffle relations~\cite{JBSH}. The latter
sums are related to one { Mellin} transform for each index set $(-1,1,1,1), 
(-1,-1,-1,1), (-1,-1,1,1), (-2,2), (-3,1), (3,1),$ $(-3,-1), (3,-1), (-1,-1,-2),
(-1,-1,2), (1,1,-2)$ and $(1,1,2)$ and to two { Mellin} transforms for the
sets $(-1,1,-2)$ and $(-1,1,2)$. Various of these harmonic sums contain 
an index $\{-1\}$. Since sums of this type do not occur in the known physics 
applications we will discuss sums without this index only from here on. 
After the algebraic relations are exploited, the seven sums with 
index sets $(2,1,1), (-2,1,1), (3,1), (-3,1), (-2,2), (4),$ and $(-4)$  
remain at this weight. 

Unlike the case for weight {\sf w = 2}, the harmonic sums with 
symmetric index pattern at weight {\sf w = 4}, $S_{-2,-2}(N)$ and 
$S_{2,2}(N)$
relate two of the above { Mellin} transforms. They have the
representation,~\cite{HS2}.
\begin{eqnarray}
\label{eq4a}
S_{-2,-2}(N) &=& \Mvec\left[\left(\frac{\ln(x)\Li_2(-x)
-2\Li_3(-x)}{x-1}\right)_+\right](N)
\nonumber
\\ 
& &+ 
\frac{\zeta_2}{2}\left[S_2(N)-S_{-2}(N)\right]-\frac{3}{2} \zeta_3 S_1(N)
\nonumber\\
&=& \frac{1}{2}\left\{\left[\beta'(N+1)+(-1)^{N} \frac{1}{2} 
\zeta_2\right]^2 - \frac{1}{6} \psi^{(3)}(N+1) +\frac{2}{5} 
\zeta_2^2\right\}
\\
\label{eq4b}
S_{2,2}(N) &=& \Mvec\left[\left(\frac{\ln(x)\Li_2(x) 
 - 2 \Li_3(x)}{x-1}\right)_+\right](N) 
+ 2 \zeta_3 S_1(N)\nonumber\\ 
&=& \frac{1}{2}\left\{\left[\psi'(N+1)-\zeta_2\right]^2-\frac{1}{6} 
\psi^{(3)}(N+1) + \frac{2}{5} \zeta_2^2\right\}~.
\end{eqnarray}
Since the { Mellin} transform of the distribution 
$[\Li_2(-x)/(x-1)]_+$, (\ref{eq55}), depends on $F_{1,2,4}(N)$ and 
$\Mvec\left[\left(\Li_2(x)/(x-1)\right)_+\right](N)$ is a basic function 
their
derivative w.r.t. $N$ is known and the { Mellin} transforms of
$[\Li_3(-x)/(x-1)]_+$ and $[\Li_3(x)/(x-1)]_+$ can be expressed by 
(\ref{eq4a},\ref{eq4b}).

One may use the relation
\begin{eqnarray}
S_{3,-1}(N) + S_{-1,3}(N) = 
S_{-1}(N) S_3(N) + S_{-4}(N)
\end{eqnarray}
to express $\Mvec\left[\Li_3(-x)/(1+x)\right](N)$  
\begin{eqnarray}
\label{eqLi3A}
\Mvec\left[\frac{\Li_3(-x)}{1+x}\right](N)
&=& 
- \Mvec\left[\frac{\Li_3(x)}{1+x}\right](N)
+ \Mvec\left[\frac{\ln(x) \Li_2(x) +\ln^2(x) \ln(1-x)/2}{1+x}\right](N)
\nonumber\\ &&
+(-1)^{N}\Biggl\{\frac{3}{5}\zeta_2^2-\frac{5}{2} \zeta_3 \ln(2) +
S_{-1}(N) S_3(N) + S_{-4}(N)
\nonumber\\ &&- 
\ln(2)\left[S_{-3}(N)-S_3(N)\right] 
+ \frac{1}{2} \zeta_2 S_{-2}(N) - \frac{7}{4} \zeta_3 S_{-1}(N)\Biggr\}
\nonumber\\ &&
- (S_{-1}(N)+\ln(2)) \zeta_3~.
\end{eqnarray}
Furthermore, the relation
\begin{eqnarray}
\frac{1}{4}
\Mvec\left[\left(\frac{\Li_3(x)}{1-x}\right)_+\right]\left(\frac{N}{2}\right)
&=& \Mvec\left[\left(\frac{\Li_3(x)}{1-x}\right)_+\right](N) +
\Mvec\left[\frac{\Li_3(x)}{1+x}\right](N)
\\ & &+
\Mvec\left[\left(\frac{\Li_3(-x)}{1-x}\right)_+\right](N)
+
\Mvec\left[\frac{\Li_3(-x)}{1+x}\right](N)
\nonumber\\ & &
+\frac{29}{40} \zeta_2^2 - 2 \Li_4\left(\frac{1}{2}\right) - \frac{3}{2}
\ln(2) \zeta_3 +\frac{1}{2} \zeta_2 \ln^2(2) - \frac{1}{12} \ln^4(2)
\nonumber
\end{eqnarray}
holds. However, similar to (\ref{eqA}), it does not lead to a  further
reduction
since $\Mvec\left[{\Li_3(-x)}/(1+x)\right](N)$ and
$\Mvec\left[{\Li_3(x)}/(1+x)\right](N)$ enter in the same way as
in (\ref{eqLi3A}).

Square argument relations may be considered for more general Nielsen integrals
as well. For $\Sf(x)/(1-x)$ one obtains the relation
\begin{eqnarray}
\label{eqS12rel}
\frac{1}{2}\Mvec\left[\left(\frac{\Sf(x)}{1-x}\right)_+\right]\left(\frac{N-1}{2}\right)
&=& \Mvec\left[\left(\frac{\Sf(x)}{1-x}\right)_+\right](N) 
+ \Mvec\left[\left(\frac{\Sf(x)}{1+x}\right)_+\right](N)
\nonumber\\ &&
+ \Mvec\left[\left(\frac{\Sf(-x)}{1-x}\right)_+\right](N) 
+ \Mvec\left[\left(\frac{\Sf(-x)}{1+x}\right)_+\right](N) 
\nonumber\\ &&
+ \Mvec\left[\left(\frac{I_1(x)}{1-x}\right)_+\right](N)
+ \Mvec\left[\left(\frac{I_1(x)}{1+x}\right)_+\right](N)
\nonumber\\ &&
+ \int_0^1 dx \frac{S_{1,2}(x^2)}{1+x}~,
\end{eqnarray}
which refers to the functions $\Sf(\pm x)/(1 \pm x)$ and $I_1(x)/(1 \pm x)$. 
$I_1(x)$ is given by, cf.~\cite{JB2},
\begin{eqnarray}
I_1(x) &=& \frac{1}{2} S_{1,2}(x^2) - S_{1,2}(x) - S_{1,2}(-x)~. 
\end{eqnarray}
The function $I_1(x)/(1-x)$ does not occur up to {\sf w = 6} for single scale 
quantities. Therefore, we do not consider square argument relations beyond 
those 
for $\Li_n(x)$ in the following.

At weight {\sf w = 4} the following basic functions contribute~:
\begin{eqnarray}
F_{6a}(N) &:=& \Mvec\left[\left(\frac{\Li_3(x)}{1-x}\right)_+\right](N)\\
F_{6b}(N) &:=& \Mvec\left[\frac{\Li_3(x)}{1+x}\right](N)\\
F_7(N) &:=& \Mvec\left[\left(\frac{S_{1,2}(x)}{x-1}\right)_+\right](N)\\
F_8(N) &:=& \Mvec\left[\frac{S_{1,2}(x)}{1+x}\right](N)~.
\end{eqnarray}
The number of basic { Mellin} transform occurring in physical processes
turns out to be even smaller. In the case of the one--loop anomalous dimensions and
the Wilson coefficients in first order in the coupling constant, at most
single harmonic sums contribute. In the case of the two--loop anomalous dimensions
$F_5(N)$ emerges as the first non--trivial function, cf.~\cite{HS1}.
The case of the 2--loop coefficient functions has been analyzed systematically
in Refs.~\cite{COHS}  for all Wilson coefficients in polarized and 
unpolarized
deeply inelastic scattering, the polarized and unpolarized { Drell--Yan}
process and the fragmentation functions, as well as the coefficient functions
for hadronic Higgs-- and pseudo--scalar Higgs--production. In particular
{ Mellin} transforms being associated to harmonic sums with an index set 
formed of chains of the elements $-1$ and $+1$ do not contribute
at this level. This does not exclude that functions of this type do not
emerge in higher orders.

\section{Sums of weight five} 
%

\vspace{1mm}
\noindent
We will consider the sums of different depth separately and assume that
the algebraic relations have been exploited already.
\subsection{Twofold Sums}             
%

\vspace{1mm}
\noindent
The following sums contribute~: $S_{\pm 4,1}(N), S_{\pm 3, \pm 2}(N)$.
We obtain the representations
\begin{eqnarray}
\label{eq41}
S_{4,1}(N) &=& -\Mvec\left[\left(\frac{\Li_4(x)}{x -1}\right)_+\right](N)
+ S_1(N) \zeta_4 - S_2(N) \zeta_3 + S_3(N) \zeta_2 
\\
\label{eqm41}
S_{-4,1}(N) &=& (-1)^{N+1} \Mvec\left[\frac{\Li_4(x)}{x 
+1}\right](N) +S_{-1}(N) \zeta_4 - S_{-2}(N) \zeta_3
+ S_{-3}(N) \zeta_2 
\nonumber\\ & & + \zeta_4 \ln(2) + \frac{3}{4} \zeta_2 \zeta_3 - \frac{59}{32} \zeta_5
\\
S_{2,3}(N) &=&  \Mvec\left[\left(
\frac{\ln(x) \left[S_{1,2}(1-x)-\zeta_3\right] + 3
\left[S_{1,3}(1-x)-\zeta_4\right]}{x-1}\right)_+\right](N)
\nonumber\\ & &
+3 \zeta_4 S_1(N)
\end{eqnarray}\begin{eqnarray}
\\%
S_{-2,3}(N) &=& 
(-1)^{N} \Mvec\left[
     \frac{
     \ln(x)\left[S_{1,2}(1-x) - \zeta_3\right] 
   + 3\left[S_{1,3}(1-x)- \zeta_4\right]
     }{1+x}\right](N)
\nonumber\\ & &
+ 3 \zeta_4 S_{-1}(N) + \frac{21}{32} \zeta_5 + 3 \zeta_4 \ln(2)
- \frac{3}{4} \zeta_2 \zeta_3
\\
S_{2,-3}(N) &=& 
(-1)^{N+1} \Mvec\left[\frac{1}{1+x}\left[\frac{1}{2}\ln^2(x) \Li_2(-x) - 2 
\ln(x) \Li_3(-x) + 3 
\Li_4(-x)\right]\right](N)\nonumber\\
& & +\frac{3}{4} \zeta_3 \left[S_{-2}(N)-S_{2}(N) \right] - \frac{21}{8} \zeta_4 
S_{-1}(N)
- \frac{41}{32} \zeta_5 - \frac{21}{8} \zeta_4 \ln(2) + \zeta_2 \zeta_3 
\\
S_{-2,-3}(N) &=& -\Mvec\left\{\left[\frac{1}{x-1}\left(\frac{1}{2}\ln^2(x) 
\Li_2(-x) -2 \ln(x) 
\Li_3(-x) + 3 \Li_4(-x)\right)\right]_+\right\}(N)\nonumber\\ & &
+ \frac{3}{4} \zeta_3 \left[S_{2}(N) - S_{-2}(N) \right] - \frac{21}{8} 
\zeta_4 S_1(N)~. 
\end{eqnarray}
In the above integrands we use the relations
\begin{eqnarray}
S_{1,2}(1-x) &=& - \Li_3(x) + \ln(x) \Li_2(x) + \frac{1}{2} \ln(1-x)
\ln^2(x) + \zeta_3 
\\
S_{1,3}(1-x) &=& - \Li_4(x) + \ln(x) \Li_3(x) - \frac{1}{2} \ln^2(x)
\Li_2(x) - \frac{1}{6} \ln^3(x) \ln(1-x) + \zeta_4~,
\end{eqnarray}
which reduce the representation. Since all Mellin transforms
\begin{eqnarray}
\frac{\Li_k(\pm x)}{x \pm 1},~~~k \leq 3,
\end{eqnarray}
are known,
the only new ones are
those of the functions
\begin{eqnarray}
\frac{\Li_4(\pm x)}{x \pm 1}
\end{eqnarray}
and the other terms follow by differentiation. Due to
\begin{eqnarray}
\label{eq1m4}
S_{1,-4}(N) &=& (-1)^{N} \Mvec\left[\frac{\Li_4(-x)
-\ln(x) \Li_3(-x)+\ln^2(x) \Li_2(-x)/2 +\ln^3(x) \ln(1+x)/6}{x 
+1}\right](N) \nonumber\\ & &
+ \frac{7}{8} \zeta_4\left[S_{-1}(N)- S_1(N)\right]
- \frac{1}{2} \zeta_2 \zeta_3 +\frac{7}{8} \zeta_4 \ln(2) + \frac{29}{32} 
\zeta_5
\\
\label{eq14}
S_{1,4}(N) &=& - \Mvec\left[\frac{S_{1,3}(1-x)}{x-1}\right] 
+ \frac{2}{5} \zeta_2^2 S_1(N) + \zeta_2 \zeta_3 - 2 \zeta_5 
\end{eqnarray}
Eqs.~(\ref{eqm41}, \ref{eq1m4}) allow to express
\begin{eqnarray}
\Mvec\left[\frac{\Li_4(-x)}{1+x}\right](N)~,
\end{eqnarray}
but (\ref{eq14}) does not lead to a new relation.

In some of the harmonic sums Mellin transforms of the functions
\begin{eqnarray}
\frac{\Li_k(-x)}{x \pm 1}
\end{eqnarray}
emerge. For odd values of $k = 2 l+1$ the harmonic sums $S_{1,-(k-1)}(N), 
S_{-(k-1),1}(N)$ and $S_{-l,-l}(N)$ allow to substitute the Mellin 
transforms of these functions in terms of Mellin transforms of basic 
functions and derivatives thereof. 
For even values of $k$ this argument applies to $\Mvec[\Li_k(-x)/(1+x)](N)$ 
but not to $\Mvec[\Li_k(-x)/(1+x)](N)$. In the latter case one may use 
relation (\ref{LIREL}).
Since in massless quantum field-theoretic calculations both denominators 
occur, one may use this decomposition based on the first two cyclotomic 
polynomials, cf. \cite{CYCL}, and the decomposition relation for 
$\Li_k(x^2)$. The corresponding Mellin transforms also require 
half--integer 
arguments. The relation
\begin{eqnarray}
\label{eqREL1}
\frac{1}{2^{k-1}} 
\Mvec\left[\left(\frac{\Li_k(x^2)}{x^2-1}\right)_+\right]
\left(\frac{N-1}{2}\right)
&=& 
 \Mvec\left[\left(\frac{\Li_k(x)}{x-1}\right)_+\right](N)
+\Mvec\left[\left(\frac{\Li_k(x)}{x+1}\right)_+\right](N)
\nonumber\\ & &
+\Mvec\left[\left(\frac{\Li_k(-x)}{x-1}\right)_+\right](N)
+\Mvec\left[\left(\frac{\Li_k(-x)}{x+1}\right)_+\right](N)
\nonumber\\ & &
- \int_0^1 dx~ \frac{\Li_k(x^2)}{1+x}
\end{eqnarray}
determines $\Mvec[\Li_k(-x)/(1+x)](N)$. For $k = 4$  the last integral in
(\ref{eqREL1}) is given by
\begin{eqnarray}
\int_0^1 dx~ \frac{\Li_4(x^2)}{1+x} &=& \frac{2}{5} \ln(2) \zeta_2^2 + 3 
\zeta_2 \zeta_3 - \frac{25}{4} \zeta_5
\end{eqnarray}
and one obtains
\begin{eqnarray}
\Mvec\left[\frac{\Li_4(-x)}{x+1}\right](N) &=&  
- \frac{1}{8} 
\Mvec\left[\left(\frac{\Li_4(x)}{x-1}\right)_+\right]\left(\frac{N-1}{2}\right)
+\Mvec\left[\left(\frac{\Li_4(x)}{x-1}\right)_+\right]\left( N \right)
\nonumber\\ & &
+\Mvec\left[\left(\frac{\Li_4(-x)}{x-1}\right)_+\right]\left( N \right)
-\Mvec\left[\frac{\Li_4(x)}{x+1}\right]\left( N \right)
\nonumber\\ & &
-\frac{1}{20} \zeta_2^2 \ln(2) - \frac{3}{8} \zeta_2 \zeta_3
+\frac{25}{32} \zeta_5~.
\nonumber\\
\end{eqnarray}
The new functions are~:
\begin{eqnarray}
F_9(N) &=& \Mvec\left[\left(
\frac{\Li_4(x)}{x - 1}\right)_+\right](N)\\
F_{10}(N) &=& \Mvec\left[  
\frac{\Li_4(x)}{1 + x}\right](N)~.
\end{eqnarray}
\subsection{Threefold Sums}             
%

\vspace{1mm}
\noindent
At this depth the sums $S_{\pm 3,1,1}(N), S_{2,2,1}(N), S_{-2,-2,1}(N),
S_{-2,2,1}(N)$ and $S_{2,-2,1}(N)$ contribute.
One derives the following representations.
\begin{eqnarray}
S_{3,1,1}(N) &=& 
\Mvec\left[\left(\frac{S_{2,2}(x)}{x-1}\right)_+\right](N)
+ \zeta_3 S_2(N) - \frac{\zeta_4}{4} S_1(N)
\\
S_{-3,1,1}(N) &=& (-1)^N \Mvec\left[\frac{S_{2,2}(x)}{1+x}\right](N)
+ \zeta_3 S_{-2}(N)-\frac{\zeta_4}{4} S_{-1}(N) \nonumber\\
& & +\frac{7}{8} \zeta_3 \zeta_2 -\frac{1}{4} \zeta_4 \ln(2) 
-\frac{7}{8} \zeta_3 \ln^2(2) + \frac{1}{3} \zeta_2 \ln^3(2) 
\nonumber\\ & &
+ \frac{15}{32} \zeta_5 -2 \ln(2) \Li_4\left(\frac{1}{2}\right) 
- 2 \Li_5\left(\frac{1}{2}\right) -\frac{1}{15} \ln^5(2) 
\\
S_{2,2,1}(N) &=& 
- \Mvec\left[\left(\frac{2 S_{2,2}(x) - 
\Li_2^2(x)/2}{x-1}\right)_+\right](N) + \zeta_2 S_{2,1}(N) - \frac{3}{10} 
\zeta_2^2 S_1(N) 
\\
S_{-2,1,-2}(N) &=& 
\Mvec\left[\left(\frac{\ln(x) S_{1,2}(-x) - 
\Li_2^2(-x)/2}{x-1}
\right)_+ \right](N) 
\nonumber\\ & &
- \frac{1}{2} \zeta_2 \left[S_{-2,1}(N)-S_{-2,-1}(N) \right]
- \left[\frac{1}{8} \zeta_3 - \frac{1}{2} \ln(2) 
\zeta_2\right] \left[S_{-2}(N) - S_2(N)\right] \nonumber\\
& & +\frac{1}{8} \zeta_2^2 S_1(N) 
\\
S_{-2,2,1}(N) &=& 
(-1)^{N+1} \Mvec\left[\frac{2 S_{2,2}(x) -
\Li_2^2(x)/2}{1+x}\right](N) + \zeta_2 S_{-2,1}(N) - \frac{3}{10}
\zeta_2^2 S_{-1}(N)\nonumber\\
& &
+ 4 \Li_5\left(\frac{1}{2}\right) 
+ 4 \Li_5\left(\frac{1}{2}\right) \ln(2)
- \frac{89}{64} \zeta_5
- \frac{9}{8} \zeta_2 \zeta_3
+ \frac{2}{15} \ln^5(2)
- \frac{2}{3} \zeta_2 \ln^3(2)
\nonumber\\ & &
+ \frac{7}{4} \zeta_3 \ln^2(2)
- \frac{3}{10} \zeta_2^2 \ln(2)
\end{eqnarray}\begin{eqnarray}
%
S_{2,1,-2}(N) &=& (-1)^{N} \Mvec\left[\frac{ 
\ln(x) S_{1,2}(-x) - \Li_2^2(-x)/2}{1+x}\right](N)
\nonumber\\ & &
- \frac{1}{2} \zeta_2 \left[S_{2,1}(N) - S_{2,-1}(N)\right]
- \left[\frac{1}{8} \zeta_3 - \frac{1}{2} \zeta_2 \ln(2) \right] 
\left[S_{2}(N) - S_{-2}(N)\right] 
\nonumber\\ & &
+ \frac{1}{8} \zeta_2^2 S_{-1}(N)
- \frac{177}{64} \zeta_5 + \frac{11}{8} \zeta_2 \zeta_3 + \frac{1}{8} 
\zeta_2^2 \ln(2)~. 
\end{eqnarray}
Here also the sums $S_{\pm 2,-1}(N)$ emerge, which is a consequence of the
representation chosen. The associated Mellin transforms are accounted for, 
however,
by the {\sf w = 3} basic functions, cf.~\cite{HS2}.
In the threefold sums the basic functions

\begin{eqnarray}
F_{11}(N)  &=& \Mvec\left[\left(\frac{S_{2,2}(x)}{x-1}\right)_+\right](N)
\\
F_{12}(N)  &=& \Mvec\left[\frac{S_{2,2}(x)}{1+x}\right](N)
\\
F_{13}(N)  &=& \Mvec\left[\left(\frac{\Li_2^2(x)}{x-1}\right)_+\right](N)
\\
F_{14}(N)  &=& \Mvec\left[\frac{\Li_2^2(x)}{1+x}\right](N)
\\
F_{15}(N)  &=& 
\Mvec\left[\frac{\ln(x) 
S_{1,2}(-x)-\Li_2^2(-x)/2+ \zeta_2^2/8}{x-1}\right](N) \\
F_{16}(N)  &=& 
\Mvec\left[\frac{\ln(x) S_{1,2}(-x)-\Li_2^2(-x)/2}{1+x}\right](N)
\end{eqnarray}
emerge.
\subsection{Fourfold Sums}             
%

\vspace{1mm}
\noindent
Here the two sums $S_{\pm 2,1,1,1}(N)$ contribute. They obey the 
representation
\begin{eqnarray}
S_{2,1,1,1}(N) &=& 
- \Mvec\left[\left(\frac{S_{1,3}(x)}{x-1}\right)_+\right](N)
+ \zeta_4 S_1(N)
\end{eqnarray}\begin{eqnarray}
S_{-2,1,1,1}(N) &=& (-1)^{N+1} \Mvec\left[\frac{S_{1,3}(x)}{1+x}\right](N)
+ \zeta_4 S_{-1}(N) + \zeta_4 \ln(2) -\frac{7}{16} \zeta_2 \zeta_3 
-\frac{1}{6} \zeta_2 \ln^3(2) \nonumber\\ & &
+\frac{7}{16} \zeta_3 \ln^2(2) 
         -\frac{27}{32} 
\zeta_5+\ln(2) \Li_4\left(\frac{1}{2}\right)+\frac{1}{30} 
\ln^5(2)+\Li_5\left(\frac{1}{2}\right)~.
\end{eqnarray}
The final basic functions contributing up to weight {\sf w = 5} are thus
\begin{eqnarray}
F_{17}(N) &=& \Mvec\left[\left(\frac{S_{1,3}(x)}{x - 1}\right)_+\right](N)\\
F_{18}(N) &=& \Mvec\left[\frac{S_{1,3}(x)}{1+x}\right](N)~.
\end{eqnarray}

Let us summarize the basic functions, which contribute up to the level of
the 3--loop anomalous dimensions, which are {\sf w = 5} quantities. They are given by
\begin{alignat}{2}
\label{eqBAS1}
{\sf w=1:}~~~& 1/(x-1)
\\
{\sf w=2:}~~~ & \ln(1+x)/(x+1)
\\
{\sf w=3:}~~~ & \Li_2(x)/(x \pm 1)
\\
{\sf w=4:}~~~ & \Li_3(x)/(x + 1), \hspace*{7mm} S_{1,2}(x)/(x \pm 1)
\\
\label{eqBAS2}
{\sf w=5:}~~~ & \Li_4(x)/(x \pm 1), \hspace*{7mm}  S_{1,3}(x)/(x + 1),
\hspace*{7mm}
              S_{2,2}(x)/(x \pm 1), \nonumber\\ &
 \Li_2^2(x)/(x + 1), \hspace*{7mm}
               [\ln(x) S_{1,2}(-x) - \Li^2_2(-x)/2]/(x \pm 1)~.
\end{alignat}
The five-fold sum $S_{1,1,1,1,1}(N)$ decomposes algebraically into a polynomial out of
single sums, \cite{HS2}.
The representation of the regularized Mellin transforms of the basic functions for 
complex argument is given in Appendix~B. 

\section{Conclusions}
%

\vspace{1mm}
\noindent
We investigated the relations between harmonic sums, which are related to
their functional value. They extend the set of algebraic relations being
implied by their quasi-shuffle property related to the index pattern. In 
quantum field theoretic calculations of massless single scale quantities up to
3--loop order and for massive quantities to 2--loop order only harmonic sums
with indices $i \neq -1$ occur. We derived the structural relations for 
this sub-algebra up to weight {\sf w = 5}. These are fractional argument-, 
integration-by-parts- and differentiation relations. The set of 69 harmonic 
sums is reduced to 30 sums by the algebraic relations. The structural 
relations imply a further reduction to 15 basic harmonic sums.
In physical applications these sums have to be known for complex arguments.
We showed that these functions can be analytically continued to meromorphic 
functions with poles at the non-positive integers referring to their 
representation in terms of Mellin-integrals. The basic sums obey recurrence 
relations for complex arguments given by corresponding harmonic sums of lower 
weight. Separating divergent contributions for $|N| \rightarrow \infty$, 
$\propto S_1^k(N),~k \geq 0, k \in {\bf N}$ algebraically, the asymptotic 
representation of the harmonic sums was derived in analytic form. The recursion 
relations and asymptotic representations allow to calculate the basic 
functions for complex arguments analytically. In the physical applications to 
{\sf w = 5}, 15 basic functions contribute, as the example of the 3--loop 
anomalous dimensions shows. This set is extended to 37 functions at {\sf w = 6} 
in case of the 3--loop Wilson coefficients, cf. \cite{JB08}. Here, 20 functions
emerge for {\sf w = 6} and two {\sf w = 5} functions 
$\Mvec[(\Li_2^2(x)/(x-1))_+](N)$ and $\Mvec[(S_{1,3}(x)/(x-1))_+](N)$ do also occur.
They do not contribute to the 3-loop anomalous dimensions. We included their 
representation in the present paper. The investigation of a large number of massless 
and massive 2--loop problems showed that the basic functions are universal up to 
isomorphies. This is expected also for the 3--loop case.

\vspace{3mm}\noindent
{\bf Acknowledgment.}\\
I would like to thank J. Ablinger, M. Kauers, C. Schneider, and J. Vermaseren for useful discussions.
This paper was supported in part by DFG Sonderforschungsbereich Transregio~9, Computergest\"utzte 
Theoretische Physik, and by the European Commission MRTN HEPTOOLS under Contract No. 
MRTN-CT-2006-035505.

\newpage
\appendix

\section{Algebraic Relations}
%

\vspace{1mm}
\noindent
In this Appendix we list the algebraic relations w.r.t. the basis derived in
the present paper. As the two-fold sums obey the simple relation (\ref{Eqeul})
we list the sums of higher depth only.
In the following we give the explicit representation of the 
harmonic sums without an index $\{-1\}$, which contribute
in physical single--scale processes up to {\sf w = 5}.
The single sums are given in (\ref{sing1}, \ref{sing2}). In the following we drop the 
common argument $N$.

\vspace{2mm}\noindent
{\underline{\sf Weight 2 Sums~:}}
\begin{eqnarray}
S_{1,1}     &=& \frac{1}{2} \left[S_1^2+S_2\right] 
\end{eqnarray}

\vspace{2mm}\noindent
{\underline{\sf Weight 3 Sums~:}}
\begin{eqnarray}
S_{1,1,1}   &=& 
\frac{1}{6} S_1^3 + \frac{1}{2} S_1 S_2 + \frac{1}{3} S_3
\end{eqnarray}

\vspace{2mm}\noindent
{\underline{\sf Weight 4 Sums~:}}
\begin{eqnarray}
S_{1,-2,1}  &=& -2 S_{-2,1,1} +S_1 S_{-2,1} 
+ S_{-3,1} + S_{-2,2}
\\ 
S_{1,1,-2} &=& 
S_{-2,1,1}-S_1 S_{-2,1}-S_{-3,1}-S_{-2,2} + S_1 S_{-3} + S_{-4}
+ \frac{1}{2} S_{-2} \left[S_1^2 +S_2\right]
\nonumber\\
\\
S_{1,2,1}   &=& 
-2 S_{2,1,1} +S_1 S_{2,1} + S_{3,1} + S_{2,2}
\\
S_{1,1,2}  &=& S_{2,1,1} 
-S_1 S_{2,1} - S_{3,1} - S_{2,2} + S_3 S_1 + S_4 + \frac{1}{2} S_2 
\left[S_1^2+S_2\right]
\\ 
S_{1,1,1,1} 
  &=& \frac{1}{4} S_{4} + \frac{1}{8} S_2^2 + \frac{1}{3} S_1 S_3 + \frac{1}{4} S_1^2 S_2 
  + \frac{1}{24} S_1^4
\nonumber\\   
\end{eqnarray}

\vspace{2mm}\noindent
{\underline{\sf Weight 5 Sums~:}}
\begin{eqnarray}  
S_{1,-3,1} &=&  -2 S_{-3,1,1} + S_1 S_{-3,1} + S_{-4,1} + S_{-3,2}
\\
S_{1,1,-3} &=& S_{-3,1,1} - S_1 S_{-3,1} - S_{-4,1} - S_{-3,2} + S_1 S_{-4} + S_{-5}
+ \frac{1}{2} S_{-3} \left[S_1^2 + S_2 \right]
\\
S_{1,3,1} &=& -2 S_{3,1,1} + S_1 S_{3,1} + S_{4,1} + S_{3,2}
\\
S_{1,1,3}   &=& S_{3,1,1} - S_1 S_{3,1} - S_{4,1} - S_{3,2} + S_1 S_4 + S_5
+ \frac{1}{2} S_3 \left[S_1^2 + S_2\right]
\\
S_{-2,-2,1} &=& \frac{1}{2}\left[-S_{-2,1,-2} +S_{-2} S_{-2,1}
+S_{4,1} + S_{-2,-3} \right] 
\\
S_{1,-2,-2} &=& \frac{1}{2} \left[-S_{-2,1,-2} +S_{-2} S_{1,-2} 
+ S_{-3,-2}+S_{1,4} \right]
\\
S_{2,-2,1} &=& 
-S_{2,1,-2}-S_{-2,2,1}+S_{3,-2}-S_{1,-4}-S_{-2} S_{1,2}+S_1 S_{2,-2}
+S_{2,-3}+S_1 S_{-2,2}+S_{-2,3} 
\nonumber\\
\\
S_{-2,1,2} &=& S_2 S_{-2,1} + S_{-4,1} + S_{2,1,-2} - S_{3,-2} + S_{1,-4} 
+ S_{-2} S_{1,2} - S_1 S_{2,-2} - S_{2,-3} 
- S_1 S_{-2,2}
\nonumber\\ &&
\end{eqnarray}\begin{eqnarray}
S_{1,-2,2} &=& -S_2 S_{-2,1} - S_{-4,1} - S_{2,1,-2} - S_{-2,2,1} 
+ S_{3,-2} 
\nonumber\\ &&
- S_{1,-4} - S_{-2} S_{1,2} + S_1 S_{2,-2} + S_{2,-3}
+ 2 S_1 S_{-2,2}+ S_{-2,3} + S_{-3,2} 
\\
S_{1,2,-2} &=& 
S_{-2} S_{1,2}+S_{1,-4}-S_1 S_{-2,2} -S_{-2,3}+S_{-2,2,1} 
\\
S_{2,1,2} &=& -2 S_{2,2,1} + S_2 S_{2,1} + S_{4,1} + S_{2,3} 
\\
S_{1,2,2} &=& S_{2,2,1}+\frac{1}{2} \left[S_2 S_{1,2}+S_{3,2}+S_{1,4}
             -S_2 S_{2,1}-S_{4,1}-S_{2,3}\right] 
\\
S_{1,-2,1,1} &=& 
-3 S_{-2,1,1,1}+S_1 S_{-2,1,1}+S_{-3,1,1}+S_{-2,2,1}+S_{-2,1,2}
\\
S_{1,1,-2,1} &=& 
3 S_{-2,1,1,1}-S_1 S_{-2,1,1}-S_{-3,1,1}-S_{-2,2,1}-S_{-2,1,2} 
\nonumber\\ & &
+ \frac{1}{2} \left[S_1 S_{1,-2,1}+S_{2,-2,1}+S_{1,-3,1}+S_{1,-2,2}\right]
\\
S_{1,1,1,-2} &=&       - S_{-2,1,1,1}
+\frac{1}{3} S_1 
\left(S_{1,1,-2}+S_{-2,1,1}-\frac{1}{2} S_{1,-2,1}\right)
\nonumber\\ && 
+ \frac{1}{3}\left(S_{2,1,-2} + S_{1,2,-2} + S_{1,1,-3}  + S_{-3,1,1}
      + S_{-2,2,1}+ S_{-2,1,2}\right)  
\nonumber\\ &&
-\frac{1}{6} \left(S_{2,-2,1} + S_{1,-3,1} + S_{1,-2,2}\right)
\\
S_{1,2,1,1} &=& -3 S_{2,1,1,1} +S_1 S_{2,1,1}+S_{3,1,1}+S_{2,2,1}+S_{2,1,2}
\\
S_{1,1,2,1} &=& 3 S_{2,1,1,1} + S_1 \left[\frac{1}{2} S_{1,2,1} - S_{2,1,1} \right]
+ \frac{1}{2} \left[S_{1,3,1} + S_{1,2,2} - S_{2,2,1} \right] - S_{3,1,1} - S_{2,1,2}
\nonumber\\
\\
S_{1,1,1,2} &=& 
-S_{2,1,1,1}+\frac{1}{3} S_1 
\left(S_{1,1,2}+S_{2,1,1}-\frac{1}{2} S_{1,2,1} \right)+\frac{1}{3} 
\Bigl[ 2S_{2,1,2} 
\nonumber\\ && 
+\frac{1}{2} S_{1,2,2}+\frac{1}{2} S_{2,2,1} +S_{1,1,3}-\frac{1}{2} 
S_{1,3,1}+S_{3,1,1} \Bigr] 
\\
S_{1,1,1,1,1}(N) &=& \frac{1}{120} S_1^5
                    +\frac{1}{12} S_2 S_1^3
                    +\frac{1}{6} S_3 S_1^2 
                    +\frac{1}{4} S_4 S_1 
                    +\frac{1}{8} S_1 S_2^2
                    +\frac{1}{6} S_2 S_3 
                    +\frac{1}{5} S_5 
\end{eqnarray}
Some of the sums of lower complexity can be further expressed using algebraic 
relations.

\newpage
\section{The Basic Functions for Complex Arguments}

\vspace{1mm}\noindent
To derive the analytic representation of the Mellin transforms of the
basic functions we first express them in terms of factorial series and
a rest part which can be decomposed algebraically.~\footnote{In
\cite{LILLE1}
asymptotic relations for non-alternating harmonic sums to low orders in
$1/N^k$ were derived. Our algorithm given below is free of these
restrictions. The main ideas were presented in January 2004 \cite{KITP}, see
also \cite{JB04}.}. We show that both parts are meromorphic
functions with poles at the non--positive integers. They obey recursion
relations in which functions of the
same class but lower complexity occur. Finally we derive the associated 
asymptotic
representations in analytic form. To make use of this formalism the basic
functions have to be mapped. The corresponding functions lay in the same 
equivalence class. 
First we remind some properties of factorial series. Then we define
the functions associated to the basic Mellin transforms and derive their
recursion relations and asymptotic representations.

\subsection{Asymptotic Representations of Factorial Series}
%

\vspace{1mm}
\noindent
The { Mellin} transforms
\begin{equation}
\Omega(z) = \int_0^1~dt~\varphi(t)~t^{z-1}
\end{equation}
with $\varphi(t)$ one of the above basic functions, suitably transformed,
is an integral representation of a factorial series
$\Omega(z)$~\cite{NIELS2,ELAND} in the complex plane $z~\in~\Cvec$. 
The basic functions are mapped such, that 
$\varphi(t)$ can be expanded into a { Taylor} series around $t=1$,
\begin{equation}
\varphi(1-t) = \sum_{k=0}^{\infty} a_k t^k~.
\end{equation}
As will be shown below, the remainder terms can be expressed in terms of 
known harmonic sums of lower complexity.

For ${\sf Re}(z) > 0$, $\Omega(z)$ is given by the factorial series
\begin{equation}
\Omega(z) = \sum_{k=0}^{\infty} a_{k+1} \frac{k!}{z(z+1) \ldots (z+k)}~.
\end{equation}
$\Omega(z)$ has poles at the
non--positive integers and one may continue $\Omega(z)$
analytically to values of $z~\in~\Cvec$ as a meromorphic function.
Let us now derive the asymptotic representation
\begin{equation}
\Omega(z) \sim \sum_{k=1}^{\infty} \frac{{\mathfrak a}_k}{z^k}~.
\end{equation}
As shown in \cite{NIELS2} the coefficients ${\mathfrak a}_k$ are obtained
by
\begin{eqnarray}
{\mathfrak a}_1 &=& a_1 \nonumber \\
{\mathfrak a}_k &=& \sum_{l=0}^{k-2} (-1)^l {\mathfrak C}_{k-l}^l
a_{k-l}~,
\end{eqnarray}
where ${\mathfrak C}_{k}^l$ are the { Stirling} numbers of 2nd kind,
\begin{eqnarray}
{\mathfrak C}_k^l &=& \frac{1}{(k-1)!} \sum_{n=0}^{k-2} (-1)^n
\binom{k-1}{n} (k-n-1)^{k+l-1}
\end{eqnarray}\begin{eqnarray}
{\mathfrak C}_k^0 &=& 1 \\
{\mathfrak C}_k^l &=& {\mathfrak C}_{k+1}^l - k {\mathfrak
C}_{k+1}^{l-1}~.
\end{eqnarray}

Not for all harmonic sums the asymptotic representation can be found in 
the way outlined above using the associated { Mellin} 
transform~\cite{HS2,JBSH}. In the case of $S_1(N)$ the function
$\psi(N+1)$ contains a logarithmic contribution
\begin{eqnarray}
\label{eq7b}
\psi(1+z)  &  =  & -\gamma_E+ \sum_{k=1}^\infty
\frac{z}{k(k+z)}\nonumber\\
           &\sim & \ln(z) + \frac{1}{2z} -\sum_{k=1}^{\infty}
                   \frac{B_{2k}}{2k~~z^{2k}}~.
\end{eqnarray}
Here $B_k$ denote the Bernoulli numbers, which can be obtained from the 
generating function \cite{BERNOU}
\begin{eqnarray}
\frac{x}{\exp(x) -1} = \sum_{k=0}^\infty B_k \frac{x^k}{k!}~.
\end{eqnarray}
Also $\psi(z)$
 and its derivatives are meromorphic functions with
poles at the non--positive integers. One may show that nested harmonic sums 
with an index pattern
\begin{eqnarray}
\{1,1, \ldots,1,a, \ldots \},~~~~~|a| > 1
\end{eqnarray}
contain a term $\propto S_1^k(N)$ as $|N| \rightarrow \infty$, where $k$ is 
the 
number of $1's$ left to $a$. These terms can always be separated algebraically
and result in a groth $\propto \ln^k(z)$ similar to the behaviour in 
(\ref{eq7b}). After this separation was  performed the remainder terms can be 
expressed by factorial series.

The function $\nu(z)$ is associated to $\psi(z)$ by
\begin{eqnarray}
\nu(z)= \ln(z) - \psi(z)
\end{eqnarray}
and removes the logarithm in (\ref{eq7b}). It obeys the
following  factorial series \cite{NIELS2} 
\begin{eqnarray}
\nu(z)= \sum_{s=0}^\infty \frac{s! \psi_s(s-1)}{z(z+1) \ldots (z+s)}~,
\end{eqnarray}
where $\psi_n(x)$ denotes the Stirling 
polynomial which obeys the recursion
\begin{eqnarray}
(x+2) \psi_n(x+1) = (x-n) \psi_n(x) + (x+1) \psi_{n-1}(x)~,
\end{eqnarray}
with the initial values
\begin{eqnarray}
\psi_0(x) = \frac{1}{2}, \hspace{1cm} \psi_1(x) = 
\frac{1}{4!} (3x+2)~.
\end{eqnarray}
The function $\beta(N)$ associated to the harmonic sum $S_{-1}(N-1)$
is represented by the factorial series
\begin{eqnarray}
\beta(z)  = \sum_{k=0}^{\infty} \frac{k!}{z(z+1) \ldots (z+k)} \cdot
\frac{1}{2^{k+1}} = \sum_{k=0}^\infty \frac{(-1)^k}{z+k}~.
\end{eqnarray}

To derive asymptotic representations for the basic function $F_i(z)|_{i=1 .. 18}$
some of the functions have to be mapped into a form, which is regular at 
$z=1$, which is indicated by a hat below. We consider the following Mellin
transforms~:
\begin{eqnarray}
    {F}_1(z) &=& \Mvec\left[\frac{\ln(1+x)}{1+x}\right](z)\\
    {F}_2(z) &=& \Mvec\left[\frac{\ln^2(1+x)-\ln^2(2)}{1-x}\right](z)\\
    {F}_3(z) &=& \Mvec\left[\frac{\ln^2(1+x)}{1+x}\right](z)\\
\hat{F}_4(z) &=& \Mvec\left[\frac{\Li_2(1-x)}{1-x}\right](z)\\
\hat{F}_5(z) &=& \Mvec\left[\frac{\Li_2(1-x)}{1+x}\right](z)
\\
\hat{F}_{6a}(z) &=& \Mvec\left[\frac{\Sf(1-x)}{1-x}\right](z)
\\
\hat{F}_{6b}(z) &=& \Mvec\left[\frac{\Sf(1-x)}{1+x}\right](z)\\
\hat{F}_7(z) &=& \Mvec\left[\frac{\Li_3(1-x)}{1-x}\right](z)\\
\hat{F}_8(z) &=& \Mvec\left[\frac{\Li_3(1-x)}{1+x}\right](z)\\
\hat{F}_{9}(z)   & =& \Mvec\left[\frac{S_{1,3}(1-x)}{1-x}\right](z)\\
\hat{F}_{10}(z) &=& \Mvec\left[\frac{S_{1,3}(1-x)}{1+x}\right](z)\\
\hat{F}_{11}(z) &=& \Mvec\left[\frac{S_{2,2}(1-x)}{1-x}\right](z)\\
\hat{F}_{12}(z) &=& \Mvec\left[\frac{S_{2,2}(1-x)}{1+x}\right](z)\\
\hat{F}_{13}(z) &=& \Mvec\left[\frac{\Li_2^2(1-x)}{1-x}\right](z)\\
\hat{F}_{14}(z) &=& \Mvec\left[\frac{\Li_2^2(1-x)}{1+x}\right](z)\\
{F}_{15}(z) &=& 
\Mvec\left[\frac{\ln(x) \Sf(-x) 
-\Li_2^2(-x)/2+\zeta_2^2/8}{1-x}\right](z)\\
{F}_{16}(z) &=& 
\Mvec\left[\frac{\ln(x) \Sf(-x) -\Li_2^2(-x)/2}{1+x}\right](z)\\
\hat{F}_{17}(z) &=& 
\Mvec\left[\frac{\Li_4(1-x)}{1-x}\right](z)
\end{eqnarray}\begin{eqnarray}
\hat{F}_{18}(z) &=& 
\Mvec\left[\frac{\Li_4(1-x)}{1+x}\right](z)~.
\end{eqnarray}

\subsection{Representation of the Basic Functions}

\vspace{1mm}\noindent
The numerator functions are related to the original basic functions, cf.~
\cite{DUDE}, by 
\begin{eqnarray}
\Li_2(x) &=& - \Li_2(1-x) - \ln(x) \ln(1-x) +\zeta_2\\
\Li_3(x) &=& -S_{1,2}(1-x) -\ln  \left( x \right) \Li_2(1-x) 
-\frac{1}{2}\,  \ln^2(x) \ln  \left( 1-x \right) 
+ \zeta_2 \ln  \left( x \right) +\zeta_3
\nonumber\\
\\   
S_{1,2}(x) &=& 
-\Li_3(1-x) +\ln(1-x) \Li_2(1-x) +\frac{1}{2}\, \ln^2(1-x)\ln(x)
+\zeta_3
\\
\Li_4(x) &=& - S_{1,3}(1-x) - \ln(x) S_{1,2}(1-x) - \frac{1}{2} \ln^2(x)
\Li_2(1-x) 
\nonumber\\ &&
- \frac{1}{6} \ln^3(x) \ln(1-x) 
+ \frac{1}{2} \ln^2(x) \zeta_2
+ \ln(x) \zeta_3 + \zeta_4
\end{eqnarray}\begin{eqnarray}
S_{1,3}(x) &=& -\Li_4(1-x) + \ln(1-x) \Li_3(1-x) - \frac{1}{2} \ln^2(1-x) 
\Li_2(1-x)
\nonumber\\ &&
- \frac{1}{6} \ln^3(1-x) \ln(x) + \zeta_4 
\\
\Li_2^2(x) &=& \Li_2^2(1-x) +2\Li_2(1-x) \ln(1-x) \ln(x) - 2 \Li_2(1-x) 
\zeta_2 
\nonumber\\ & &+ \ln^2(1-x)\ln^2(x) - 2 \ln(1-x) \ln(x) \zeta_2 +\zeta_2^2
\\
S_{2,2}(x) &=& - S_{2,2}(1-x) + \ln(1-x) S_{1,2}(1-x) +\frac{\zeta_4}{4}
\nonumber\\ & & - \left[ \Li_3(1-x) - \ln(1-x) \Li_2(1-x) - \zeta_3 \right] 
\ln(x)
+\frac{1}{4} \ln^2(1-x) \ln^2(x)~.
\end{eqnarray}

\vspace{1mm} \noindent
In the above we arranged the representations such, that the remainder 
Mellin transforms can be obtained either by algebraic relations or
differentiation of Mellin transforms being derived above or are known 
otherwise, cf.~\cite{HS2,JBHK,BM1}. We remind the following relations~:
\begin{eqnarray}
\label{pre1}
\Mvec[ \ln(1-x) \Li_2(1-x)](N) &=& \frac{2}{N+1} \Bigl[ S_1(N+1) S_2(N+1)
+ S_3(N+1) - \zeta_2 S_1(N+1) \nonumber\\ &&
- \frac{1}{2} S_{2,1}(N+1) \Bigr]~, 
\\
\Mvec[ \ln^2(1-x) \Li_2(1-x)](N) &=& - \frac{1}{N+1} \Bigl[6 S_{1,1,2}(N+1) 
+ 4 S_{1,2,1}(N+1) + 2 S_{2,1,1}(N) 
\nonumber\\ &&
-6 \zeta_2 S_{1,1}(N+1)-2 \zeta_3 S_1(N+1)\Bigr]~,
\\
\Mvec[ \ln(1-x) \Li_3(1-x)](N) &=& - \frac{1}{N+1} \Bigl[3 S_{1,1,2}(N+1) 
+  S_{1,2,1}(N+1) 
\nonumber\\ &&
- \frac{3}{2} \zeta_2 (S_1^2(N+1)+S_2(N+1)) + \zeta_3 S_1(N+1) 
\Bigr]~,
\\ 
\label{pre2}
\Mvec[ \ln(1-x) S_{1,2}(1-x)](N) &=& 
\frac{1}{N+1}\Bigl[2 S_{1,3}(N+1) + S_{3,1}(N+1) + S_{2,2}(N+1) 
\nonumber\\ &&
- \zeta_2 S_2(N+1)
- 2 \zeta_3 S_1(N+1)
\Bigr]~. 
\end{eqnarray}
By summation one obtains from (\ref{pre1}--\ref{pre2})
\begin{eqnarray}
\Mvec\left[\left(\frac{\ln(1-x) \Li_2(1-x)}{x-1}\right)_+\right](N) &=& 
2 S_{1,1,2}(N) + S_{1,2,1}(N) - 2 \zeta_2 S_{1,1}(N)
\nonumber\\ &=&
-S_1(N) S_{2,1}(N) - S_{3,1}(N) - S_{2,2}(N) 
\nonumber\\ &&
+ 2 S_3(N) S_1(N) + 2 S_4(N) 
\nonumber\\ &&
+ 2 (S_2(N) - \zeta_2) S_{1,1}(N) 
\\
\Mvec\left[\left(\frac{\ln^2(1-x) \Li_2(1-x)}{x-1}\right)_+\right](N) &=& 
-6 S_{1,1,1,2}(N) - 4 S_{1,1,2,1}(N) - 2 S_{1,2,1,1}(N)
\nonumber\\ &&
+ 6 \zeta_2 S_{1,1,1}(N) + 2 \zeta_3 S_{1,1}(N)
\\
\Mvec\left[\left(\frac{\ln(1-x) \Li_3(1-x)}{x-1}\right)_+\right](N) &=& - 3 S_{1,1,1,2}(N)
- S_{1,1,2,1}(N) + 3 \zeta_2 S_{1,1,1}(N) 
\nonumber\\ && - \zeta_3 S_{1,1}(N)
\nonumber\\ &=&
-S_1(N) S_{1,1,2}(N) + S_{1,2,2}(N) 
\nonumber\\ &&
- S_2(N) S_{1,2}(N) 
- S_{3,2}(N) - S_{1,4}(N) - S_{1,1,3}(N) 
\nonumber\\ &&
+ 3 \zeta_2 S_{1,1,1}(N) - S_{1,1}(N) \zeta_3
\\
\Mvec\left[\left(\frac{\ln(1-x) S_{1,2}(1-x)}{x-1}\right)_+\right](N) &=& 
2 S_{1,1,3}(N) + S_{1,3,1}(N) + S_{1,2,2}(N)  -\zeta_2 S_{1,2}(N) 
\nonumber\\ &&
- 2 \zeta_3 S_{1,1}(N)
\\
(-1)^N \Mvec\left[\frac{\ln(1-x) \Li_2(1-x)}{x+1}\right](N) &=& 
2 S_{-1,1,2}(N) + S_{-1,2,1}(N) - 2 \zeta_2 S_{-1,1}(N) 
\nonumber\\ &&
- \frac{7}{8} \zeta_2^2 + \frac{1}{8} \ln^4(2) + 3 \Li_4\left(\frac{1}{2}\right) + \frac{3}{4} \ln^2(2)
\\
(-1)^N \Mvec\left[\frac{\ln^2(1-x) \Li_2(1-x)}{x+1}\right](N) &=& 
-6 S_{-1,1,1,2}(N) - 4 S_{-1,1,2,1}(N) - 2 S_{-1,2,1,1}(N)
\nonumber\\ &&
+ 6 \zeta_2 S_{-1,1,1}(N) + 2 \zeta_3 S_{-1,1}(N) 
\nonumber\\
&& - \frac{21}{4} \zeta_2 \zeta_3 - \frac{1}{2} \zeta_2 \ln^3(2) + \frac{57}{20} 
\zeta_2^2 \ln(2) - \frac{1}{10} \ln^5(2) 
\nonumber\\ &&
- \frac{29}{32} \zeta_5 + 12 \Li_5\left(\frac{1}{2}\right)
\\
(-1)^N \Mvec\left[\frac{\ln(1-x) \Li_3(1-x)}{x+1}\right](N) &=& - 3 S_{-1,1,1,2}(N)
- S_{-1,1,2,1}(N) + 3 \zeta_2 S_{-1,1,1}(N) 
\nonumber\\  &&
- \zeta_3 S_{-1,1}(N) 
+ \frac{35}{16} \zeta_2 \zeta_3 + \frac{5}{12} \zeta_2 \ln^3(2) 
\nonumber\\ &&
- \frac{53}{40} \zeta_2^2 \ln(2)
+ \frac{7}{16} \zeta_3 \ln^2(2) 
- \frac{29}{64} \zeta_5 + \frac{1}{30} \ln^5(2)
\nonumber\\ &&
-4  \Li_5\left(\frac{1}{2}\right)  
\\
(-1)^N \Mvec\left[\frac{\ln(1-x) S_{1,2}(1-x)}{x+1}\right](N) &=& 
2 S_{-1,1,3}(N) + S_{-1,3,1}(N) + S_{-1,2,2}(N)  -\zeta_2 S_{-1,2}(N) 
\nonumber\\ &&
- 2 \zeta_3 S_{-1,1}(N) 
\nonumber\\ &&
- \frac{15}{8} \zeta_2 \zeta_3 + \frac{1}{6} \zeta_2 \ln^3(2) - \frac{11}{40} \zeta_2^2 \ln(2)
+ \frac{7}{8} \zeta_3 \ln^2(2) 
\nonumber
\end{eqnarray}\begin{eqnarray}
&&
+ \frac{159}{64} \zeta_5 - \frac{1}{60} \ln^5(2) + 2 
\Li_5\left(\frac{1}{2}\right)
\end{eqnarray}
The above expressions depend on harmonic sums, for which we derive
the following relations. They are used to derive
the asymptotic representations only.
\begin{eqnarray}  
S_{-1,1,1}(N) &=& S_{1,1,-1}(N) + \frac{1}{2}\Bigl[S_1(N) S_{-1,1}(N) + S_{-2,1}(N) + S_{-1,2}(N)
\nonumber\\ &&
- S_1(N) S_{1,-1}(N) - S_{2,-1}(N) - S_{1,-2}(N)\Bigr]
\\
S_{1,1,-1}(N) &=& (-1)^N \frac{1}{2}F_3(N) + \ln(2)\left[S_{1,-1}(N) - S_{1,1}(N)\right]
+ \frac{1}{2} \ln^2(2) \left[S_1(N) - S_{-1}(N)\right] 
\nonumber\\ &&
- \frac{1}{6} \ln^3(2)
\\
S_{1,1,2}(N) &=& -\hat{F}_7(N) + \zeta_2 S_{1,1}(N) - \zeta_3 S_1(N) + \frac{2}{5} \zeta_2^2
\\
S_{-1,1,2}(N) &=& (-1)^N \hat{F}_8(N) + \zeta_2 S_{-1,1}(N) - \zeta_3 S_{-1}(N)
- \Li_4\left(\frac{1}{2}\right) + \frac{9}{20} \zeta_2^2 - \frac{7}{8} \zeta_3 \ln(2)
\nonumber\\ &&
- \frac{1}{2} \zeta_2 \ln^2(2)-\frac{1}{24} \ln^4(2)
\\
S_{1,2,-1}(N) &=& (-1)^N \Mvec \left[\frac{2 S_{1,2}(-x) + \ln(1+x) \Li_2(-x)}
{1+x}\right](N) - \ln(2) \left[S_{1,2}(N) - S_{1,-2}(N)\right]
\nonumber\\ &&
- \frac{1}{2} \zeta_2 S_{1,-1}(N) + 
\left[\frac{1}{4} \zeta_3 - \frac{1}{2} \zeta_2 \ln(2)\right] \left[S_1(N) - 
S_{-1}(N)\right] + \frac{6}{5} \zeta_2^2 - 3 \Li_4\left(\frac{1}{2}\right)
\nonumber\\ &&
- \frac{23}{8} \zeta_3 \ln(2) + \zeta_2 \ln^2(2) - \frac{1}{8} \ln^4(2)
\\
S_{1,1,3}(N) &=& - \hat{F}_{12}(N) +\zeta_3 S_{1,1}(N) - \frac{\zeta_2^2}{10} S_1(N) 
+ 2 \zeta_5 - \zeta_2 \zeta_3 
\\
S_{1,2,2}(N) &=&  - \frac{1}{2} \hat{F}_{13}(N) - 2 S_{1,1,3}(N) + \zeta_2 S_{1,2}(N) + 2 \zeta_3 
S_{1,1}(N) - 
\frac{\zeta_2^2}{2}
S_1(N) + \zeta_2 \zeta_3 - \frac{3}{2} \zeta_5
\nonumber\\ 
\\
S_{-1,1,3}(N) &=& (-1)^N \hat{F}_{12}(N) + \zeta_3 S_{-1,1}(N) 
- \frac{\zeta_2^2}{10} S_{-1}(N) 
+ 2 \left[\Li_5\left(\frac{1}{2}\right) + \Li_4\left(\frac{1}{2}\right) \ln(2) \right]
\nonumber\\ && 
+ \frac{1}{15} \ln^5(2) - \frac{1}{3} \zeta_2 \ln^3(2) + \frac{3}{8} \zeta_2^2 \ln(2)
+ \frac{1}{16} \zeta_2 \zeta_3 -\frac{151}{64} \zeta_5
\\ 
S_{-1,3,1}(N) &=& S_{1,3,-1}(N) + S_1(N) S_{-1,3}(N) + S_{-1,4}(N) - S_{-1}(N) S_{1,3}(N)-S_{1,-4}(N)
\\
S_{1,3,-1}(N) &=& (-1)^{N+1} \Mvec\left[\frac{\ln(1+x) \Li_3(-x) + \Li^2_2(-x)/2}{1+x}\right](N)
+ \ln(2) \left[S_{1,-3}(N) - S_{1,3}(N)\right]
\nonumber\\ && + \left(-\frac{1}{8} \zeta_2^2 + \frac{3}{4} \zeta_3 \ln(2) \right)\left[S_1(N)
- S_{-1}(N)\right] - \frac{1}{2} \zeta_2 S_{1,-2}(N) + \frac{3}{4} \zeta_3 S_{1,-1}(N)
\nonumber\\ &&
+ \frac{1}{8} \zeta_2^2 \ln(2) + \frac{1}{2} \zeta_3 \ln^2(2) - \frac{1}{3} 
\zeta_2 \ln^3(2)
- \frac{125}{64} \zeta_5 + \frac{1}{15} \ln^5(2) \nonumber\\ &&
+ 2 \left[
\Li_5\left(\frac{1}{2}\right) + \Li_4\left(\frac{1}{2}\right) \ln(2)\right]
- \frac{1}{16} \zeta_2 \zeta_3
\\
S_{-1,2,2}(N) &=&  (-1)^N \frac{1}{2} \hat{F}_{14}(N) - 2 S_{-1,1,3}(N)+  \zeta_2 S_{-1,2}(N)
+  2 \zeta_3 S_{-1,1}(N) - \frac{\zeta_2^2}{2} S_{-1}(N) \nonumber\\ &&
- 4 \Li_5\left(\frac{1}{2}\right) + \frac{1}{30} \ln^5(2) - \frac{1}{3} \zeta_2 \ln^3(2)
- \frac{53}{40} \zeta_2^2 \ln(2) + \zeta_2 \zeta_3 + \frac{19}{8} \zeta_5
\end{eqnarray}
\begin{eqnarray}
S_{-1,1,1,2}(N) &=& -(-1)^N \hat{F}_{18}(N) + \zeta_2 S_{-1,1,1}(N)
- \zeta_3 S_{-1,1}(N) + \frac{2}{5} \zeta_2^2 S_{-1}(N) \nonumber\\
&&
- 2 \Li_5\left(\frac{1}{2}\right) - \ln(2) \Li_4\left(\frac{1}{2}\right)
- \frac{1}{40} \ln^5(2) 
+ \frac{1}{3} \zeta_2 \ln^3(2) - \frac{1}{20} \zeta_2^2 \ln(2)
\nonumber\\ &&
+ \frac{7}{8} \zeta_2 \zeta_3 - \frac{1}{32} \zeta_5
\\
S_{-1,1,2,1}(N) &=& (-1)^{N+1} \Mvec\left[\frac{3 S_{1,3}(x) + 2 S_{1,2}(x)
\ln(1-x) + \ln^2(1-x) \Li_2(x)/2}{1+x}\right](N) 
\nonumber\\
&&
+ \zeta_2 S_{-1,1,1}(N)
 - \frac{9}{16} \zeta_2 \zeta_3 - \frac{5}{12} \zeta_2 \ln^3(2) - 
\frac{19}{40} \zeta_2^2 \ln(2) + \frac{23}{16} \zeta_3 \ln^2(2) - 
\frac{23}{64} 
\zeta_5 
\nonumber\\ &&
+ \frac{13}{120} \ln^5(2) + 3 \ln(2) \Li_4\left(\frac{1}{2}\right)
+ 2 \Li_5 \left(\frac{1}{2}\right) 
\\
S_{-1,2,1,1}(N) &=& (-1)^N \Mvec\left[\frac{\ln(1-x) S_{1,2}(x) + 3 
S_{1,3}(x)}
{1+x}\right](N) + \zeta_3 S_{-1,1}(N) \nonumber\\
&& + \frac{11}{8} \zeta_2 \zeta_3 + \frac{5}{12} \zeta_2 \ln^3(2) - 
\frac{1}{8} \zeta_2^2 \ln(2) - \frac{11}{8} \zeta_3 \ln^2(2) - \frac{11}{120} 
\ln^5(2) 
\nonumber\\ &&
+\frac{81}{64} \zeta_5
- 3 \ln(2) \Li_4\left(\frac{1}{2}\right) - 
4 \Li_5\left(\frac{1}{2}\right)~.
\end{eqnarray}
The above relations require the function
\begin{eqnarray}  
\Mvec\left[\frac{\ln^3(1-x)}{1+x}\right](N)~,
\end{eqnarray}
which is related to
$\Mvec\left[\ln^3(1+x)/(1+x)\right](N)$ by
\begin{eqnarray}  
S_{-1,1,1,1}(N) &=& (-1)^{N+1} \frac{1}{6} \Mvec\left[\frac{\ln ^3(1-x)}{1+x}
\right](N) - \Li_4\left(\frac{1}{2}\right)
\\  
S_{1,1,1,-1}(N) &=& (-1)^{N+1} \frac{1}{6} \Mvec\left[\frac{\ln ^3(1+x)}{1+x}
\right](N) + \ln(2) \left[S_{1,1,-1}(N) - S_{1,1,1}(N) \right] 
\\
S_{-1,1,1,1}(N) &=& - S_{1,1,1,-1}(N) + S_1(N) \frac{1}{3} \left[ 
S_{-1,1,1}(N)
+ S_{1,1,-1}(N) - \frac{1}{2} S_{1,-1,1}(N)\right]
\nonumber\\ &&
- \frac{1}{6} \left[S_{2,-1,1}(N) + S_{1,-2,1}(N) + S_{1-1,2}(N)\right]
+ \frac{1}{3} \bigl[S_{-1,1,2}(N) + S_{-2,1,1}(N) 
\nonumber\\ &&
+ S_{-1,2,1}(N)
+ S_{1,1,-2}(N) + S_{2,1,-1}(N) + S_{1,2,-1}(N)\bigr]~.
\end{eqnarray}
The sums associated to the index sets $\{1,1,-1\}$  and $\{1,2,-1\}$ are
represented by $F_2(N)$,  resp.  $S_{1,2,-1}(N)$ and $S_{-1,1,2}(N)$, using 
the
algebraic relations \cite{JBSH}.
We use the subsidiary functions
\begin{eqnarray}  
H_1(N) &=& \Mvec\left[\frac{2 S_{1,2}(-x) + \ln(1+x) \Li_2(-x)}{1+x}\right](N)\\
H_2(N) &=& \Mvec\left[\frac{\ln(1+x) \Li_3(-x) + 
\Li^2_2(-x)/2}{1+x}\right](N)\\
H_3(N) &=& \Mvec\left[\frac{\ln^3(1+x)}{1+x}\right](N)
\end{eqnarray}
to obtain the asymptotic representations, cf. Sect.~B.4. The other sums are 
two-fold and their explicit representations  were given in \cite{HS2} and in 
the previous sections.

Changing the arguments from $z$ to $1-z$ in the numerator functions 
leads to relations like
\begin{eqnarray}  
\Mvec\left[ \frac{\Li_2(z) -\zeta_2}{z \pm 1}\right](N) &=& 
- \Mvec\left[ \frac{\Li_2(1-z)}{z \pm 1}\right](N) - \frac{d}{dN} 
\Mvec\left[\frac{\ln(1-z)}{z \pm 1}\right](N)
\\
\Mvec\left[ \frac{\Li_3(z) -\zeta_3}{z \pm 1}\right](N) &=& 
- \Mvec\left[ \frac{S_{1,2}(1-z)}{z \pm 1}\right](N) 
- \frac{d}{dN} \Mvec\left[\frac{\Li_2(1-z) + \zeta_2}{z \pm 1}\right](N)  
\nonumber\\ &&
- \frac{d^2}{dN^2} \frac{1}{2} \Mvec\left[\frac{\ln(1-z)}{z \pm 1}\right](N)  
\\
\Mvec\left[ \frac{S_{1,2}(z) -\zeta_3}{z \pm 1}\right](N) &=& 
- \Mvec\left[ \frac{\Li_3(1-z)}{z \pm 1}\right](N) 
+ \Mvec\left[\frac{\ln(1-x) \Li_2(1-z)}{z \pm 1}\right](N)  
\nonumber\\ &&
+ \frac{d}{dN} \frac{1}{2} \Mvec\left[\frac{\ln^2(1-z)}{z \pm 1}\right](N)  
\\
\Mvec\left[ \frac{\Li_4(z) -\zeta_4}{z \pm 1}\right](N) &=& 
- \Mvec\left[ \frac{S_{1,3}(1-z)}{z \pm 1}\right](N) 
- \frac{d}{dN} \Mvec\left[\frac{S_{1,2}(1-z)-\zeta_3}{z \pm 1}\right](N)  
\nonumber\\ &&
- \frac{d^2}{dN^2} \frac{1}{2} \Mvec\left[\frac{\Li_2(1-z)-\zeta_2}{z \pm 1}\right](N)  
- \frac{d^3}{dN^3} \frac{1}{6} \Mvec\left[\frac{\ln(1-z)}{z \pm 1}\right](N)  
\nonumber\\ 
\\
\Mvec\left[ \frac{S_{1,3}(z) -\zeta_4}{z \pm 1}\right](N) &=& 
- \Mvec\left[ \frac{\Li_4(1-z)}{z \pm 1}\right](N) 
- \frac{d}{dN} \frac{1}{6} \Mvec\left[\frac{\ln^3(1-z)}{z \pm 1}\right](N)  
\nonumber\\ &&
+ \Mvec\left[\frac{\ln(1-z)\Li_3(1-z)}{z \pm 1}\right](N)  
- \frac{1}{2}\Mvec\left[\frac{\ln^2(1-z)\Li_2(1-z)}{z \pm 1}\right](N)  
\nonumber\\ 
\Mvec\left[ \frac{\Li_2^2(z) -\zeta_2^2}{z \pm 1}\right](N) &=& 
 \Mvec\left[ \frac{\Li_2^2(1-z)}{z \pm 1}\right](N) 
 -2 \zeta_2 \Mvec\left[\frac{\Li_2(1-z)}{z \pm 1}\right](N)  
\nonumber\\ &&
 -2 \zeta_2 \frac{d}{dN} \Mvec\left[\frac{\ln(1-z)}{z \pm 1}\right](N)  
+ 2 \frac{d}{dN} \Mvec\left[\frac{\ln(1-z)[\Li_2(1-z)}{z \pm 1}\right](N)  
\nonumber\\ &&
+ \frac{d^2}{dN^2} \Mvec\left[\frac{\ln^2(1-z)}{z \pm 1}\right](N)  
\\
\Mvec\left[ \frac{S_{2,2}(z) -\zeta_4/4}{z \pm 1}\right](N) &=& 
 \Mvec\left[ \frac{S_{2,2}(1-z)}{z \pm 1}\right](N)
- \Mvec\left[ \frac{\ln(1-z) S_{1,2}(1-z)}{z \pm 1}\right](N) 
\nonumber\\
&&
-\frac{d}{dN}\Mvec\left[\frac{\Li_3(1-z)-\zeta_3 -\ln(1-z) \Li_2(1-z)}{z \pm 1}\right](N)  
\nonumber\\ &&
+ \frac{d^2}{dN^2} \frac{1}{4}\Mvec\left[\frac{\ln^2(1-z)}{z \pm 1}\right](N)  
\end{eqnarray}
which are used to express the asymptotic representations,~Sect.~B.4.

\subsection{Recursion Relations}
The following recursion relations hold for the basic functions~:
\begin{eqnarray}  
\label{eq7rb}
\psi(1+z)  &=& \psi(z) + \frac{1}{z}
\\ 
\psi^{(n)}(1+z) &=& \psi^{(n)}(z) + (-1)^n \frac{n!}{z^{n+1}}
\end{eqnarray}\begin{eqnarray}
\beta(z+1) &=& -\beta(z) + \frac{1}{z}
\\
\beta^{(n)}(z+1) &=& -\beta^{(n)}(z) + (-1)^n \frac{n!}{z^{n+1}}
\\
F_1(z) &=& - F_1(z-1) + \frac{1}{z}\left[\ln(2)-\beta(z+1)\right]
\\
F_2(z) &=& F_2(z-1)  +\frac{2}{z} F_1(z) 
\\
F_3(z) &=& -F_3(z-1)  +\frac{1}{z}\left[\ln^2(2) - 2
             F_1(z+1)\right]
\\
F_4(z) &=& F_4(z-1) - \frac{1}{z} \left[\zeta_2 -
\frac{S_1(z)}{z}\right]
\\
F_5(z) &=& -F_5(z-1) - \frac{1}{z} \left[\zeta_2 - 
\frac{S_1(z)}{z}\right]
\\
F_{6a}(z) &=& F_{6a}(z-1) -\frac{\zeta_3}{z} + \frac{1}{z^2} \left[\zeta_2 -
\frac{S_1(z)}{z}\right]
\\
F_{6b}(z) &=& -F_{6b}(z-1) + \frac{\zeta_3}{z} -\frac{1}{z^2} \left[\zeta_2 - 
\frac{S_1(z)}{z}\right]
\\
F_7(z) &=& F_7(z-1) +\frac{\zeta_3}{z} - \frac{1}{2 z^2} \left[
S_1^2(z)+S_2(z)\right]
\\
F_8(z) &=& - F_8(z-1) - \frac{\zeta_3}{z} + \frac{1}{2 z^2} \left[
S_1^2(z)+S_2(z)\right]
\\
F_9(z) &=& F_{9}(z-1) +\frac{\zeta_4}{z} - \frac{\zeta_3}{z^2}
+ \frac{\zeta_2}{z^3}  - \frac{1}{z^4} S_1(z)
\\ 
F_{10}(z) &=& -F_{10}(z-1) - \frac{\zeta_4}{z} + \frac{\zeta_3}{z^2}
- \frac{\zeta_2}{z^3}  + \frac{1}{z^4} S_1(z)
\\ 
F_{11}(z) &=& F_{11}(z-1) + \frac{\zeta_4}{4 z} - \frac{\zeta_3}{z^2}
+ \frac{1}{2 z^2} \left[S_1^2(z) +  S_2(z) \right] 
\\
F_{12}(z) &=& -F_{12}(z-1) + \frac{\zeta_4}{4 z} - \frac{\zeta_3}{z^2}
+ \frac{1}{2 z^2} \left[S_1^2(z) +  S_2(z) \right] 
\\
F_{13}(z) &=& F_{13}(z-1) + \frac{\zeta_2^2}{z} - \frac{4\zeta_3}{z^2} 
- \frac{2 \zeta_2}{z^2} S_1(z) + \frac{2 S_{2,1}(z)}{z^2} + \frac{2}{z^3} 
\left[S_1^2(z)+S_2(z)\right] 
\nonumber\\
\\
F_{14}(z) &=& -F_{14}(z-1) + \frac{\zeta_2^2}{z} - \frac{4\zeta_3}{z^2} 
- \frac{2 \zeta_2}{z^2} S_1(z) + \frac{2 S_{2,1}(z)}{z^2} + \frac{2}{z^3} 
\left[S_1^2(z)+S_2(z)\right] 
\nonumber\\
\\
F_{15}(z) &=& F_{15}(z-1) - R_1(z)
\\
F_{16}(z) &=& -F_{16}(z-1)+ R_1(z)
\\
F_{17}(z) &=& F_{17}(z-1) + \frac{\zeta_4}{z} - \frac{1}{6 z^2}\left[
S_1^3(z) + 3 S_1(z) S_2(z) + 2 S_3(z)\right] 
\\
F_{18}(z) &=& -F_{18}(z-1) + \frac{\zeta_4}{z} - \frac{1}{6 z^2}\left[
S_1^3(z) + 3 S_1(N) S_2(z) + 2 S_3(z)\right]~,
\end{eqnarray}
with
\begin{eqnarray}
R_1(z) &=& \Mvec\left[\ln(z) S_{1,2}(-x) - \frac{1}{2}\Li_2^2(-x) + 
\frac{\zeta_2^2}{8} \right](z)~,
\end{eqnarray}\begin{eqnarray}
\Mvec\left[\Li_2^2(-x)\right](z) &=& \frac{\zeta_2^2}{4(z+1)} - \frac{\ln(2) 
\zeta_2}{(z+1)^2} + \frac{2 
\ln^2(2)}{(z+1)^3} - \frac{4}{(z+1)^3} F_1(z+1) 
\nonumber\\ &&
- \frac{2}{(z+1)^2} \Mvec\left[\frac{\Li_2(-x)}{1+x}
\right](z+1) 
\\
\Mvec\left[\ln(x) S_{1,2}(-x)\right](z) &=& - \frac{\zeta_3}{8(z+1)^2} + \frac{\ln^2(2)}{2(z+1)^3} 
-\frac{1}{(z+1)^3} F_1(z+1) 
\nonumber\\ &&
- \frac{1}{2(z+1)} \frac{\partial}{\partial z} \left[
\frac{\ln^2(2)}{z+1} - \frac{2}{z+1} F_1(z+1)\right]~, 
\end{eqnarray}
cf. (\ref{R2P}).

The recursion relations can  be obtained from the integral--representations 
of the { Mellin} transforms. Any point in the analytic region of the basic 
functions is connected by the recursions which have to be applied until the
asymptotic region $|z| \geq z_{\rm asympt}$ with $z_{\rm asympt} \simeq 15 
... 20$ is reached.  
\subsection{Asymptotic Representations}

\vspace{1mm}\noindent
The asymptotic representations of functions contributing to the single 
harmonic sums and the basic functions $F_i(z)$ and $\hat{F}_i(z)$ are
given in the following. We maintain contributions up to $O(1/z^{19})$, to 
reach double precision accuracy in a fully {\sf analytic} representation.
Higher order terms can calculated if needed in more special numerical 
applications.
\begin{eqnarray}
\label{eq7ab}
\psi(z)  &\sim& 
\ln(z) - \frac{1}{2z} -\sum_{k=1}^{\infty} 
\frac{B_{2k}}{2k~~z^{2k}}\\
\psi^{(n)}(z) &\sim& 
(-1)^{n-1} \left[\frac{(n-1)!}{z^n} + \frac{n!}{2 
z^{n+1}} + \sum_{k=1}^{\infty} B_{2k} \frac{(2k+n-1)!}{(2k-1)!
~~z^{2k+n}}\right]\\
\psi(z) &\sim& 
\ln  \left( z \right) 
-\frac{1}{2}\,\frac{1}{z}-\frac{1}{12}\,\frac{1}{z^2}+{\frac 
{1}{120}}\,
\frac{1}{z^4}-{\frac {1}{252}}\,\frac{1}{z^6}+{\frac 
{1}{240}}\,\frac{1}{z^8}-{\frac
{1}{132}}\,\frac{1}{z^{10}}+{\frac 
{691}{32760}}\,\frac{1}{z^{12}}
\nonumber\\ &&
-\frac{1}{12}\,\frac{1}{z^{14}}
+{
\frac {3617}{8160}}\,\frac{1}{z^{16}}-{\frac 
{43867}{14364}}\,\frac{1}{z^{18}}+{
\frac {174611}{6600}}\,\frac{1}{z^{20}}+O\left(\frac{1}{z^{22}}\right)
\\
\psi'(z) &\sim& 
\frac{1}{z}+\frac{1}{2}\,\frac{1}{z^2}+\frac{1}{6}\,\frac{1}{z^3}
-\frac{1}{30}\,\frac{1}{z^5}+\frac{1}{42}\,\frac{1}{z^7}\
-\frac{1}{30}\,\frac{1}{z^9}+{\frac {5}{66}}\,\frac{1}{z^{11}}
-{\frac {691}{2730}}\,\frac{1}{z^{13}}
+\frac{7}{6}\,\frac{1}{z^{15}}
\nonumber\\ & &
-{\frac {3617}{510}}\,\frac{1}{z^{17}}
+{\frac {43867}{798}}\,\frac{1}{z^{19}}-{\frac 
{174611}{330}}\,\frac{1}{z^{21}}
+O\left(\frac{1}{z^{23}}\right)
\\
\psi^{(2)}(z) &\sim& 
-\frac{1}{z^2}-\frac{1}{z^3}-\frac{1}{2}\,\frac{1}{z^4}+\frac{1}{6} 
\frac{1}{z^6}-\frac{1}{6}\,\frac{1}{z^8}+\frac{3}{10}\,
\frac{1}{z^{10}}-\frac{5}{6}\,\frac{1}{z^{12}}+{\frac 
{691}{210}}\,\frac{1}{z^{14}}-{\frac {35}{2}}\,
\frac{1}{z^{16}}
\nonumber\\ & &
+{\frac {3617}{30}}\,\frac{1}{z^{18}}
-{\frac {43867}{42}}\,\frac{1}{z^{20}}
+O\left(\frac{1}{z^{22}}\right)
\\
\psi^{(3)}(z) &\sim& 
\frac{2}{z^3}+\frac{3}{z^4}+\frac{2}{z^5}-\frac{1}{z^7}
+\frac{4}{3}\,\frac{1}{z^9}-\frac{3}{z^{11}}
+\frac{10}{z^{13}}-{\frac 
{691}{15}}\,\frac{1}{z^{15}}+\frac{280}{z^{17}}
\nonumber\\ & &
-{\frac {10851}{5}}\,\frac{1}{z^{19}}+{\frac 
{438670}{21}}\,\frac{1}{z^{21}}
+O\left(\frac{1}{z^{23}}\right)
\\
\psi^{(4)}(z) &\sim& 
-\frac{6}{z^{4}}-\frac{12}{z^{5}}-\frac{10}{z^{6}}+\frac{7}{z^{8}}
-\frac{12}{z^{10}}+\frac{33}{z^{12}}
-\frac{130}{z^{14}}+\frac{691}{z^{16}}-\frac{4760}{z^{18}}
\nonumber\\ & &
+{\frac {206169}
{5}}\,\frac{1}{z^{20}}
+O\left(\frac{1}{z^{23}}\right)
\end{eqnarray}\begin{eqnarray}
\psi^{(5)}(z) &\sim& 
\frac{24}{z^{5}}+\frac{60}{z^6}+\frac{60}{z^7}-\frac{56}{z^9}
+\frac{120}{z^{11}}-\frac{396}{z^{13}}+\frac{1820}{z^{15}}
-\frac{11056}{z^{17}}+\frac{85680}{z^{19}}
\nonumber\\ & &
-\frac{824676}{z^{21}}
+O\left(\frac{1}{z^{23}}\right)
\\
\beta(z) 
&\sim& \frac{1}{2z} + \sum_{k=1}^\infty \frac{(-1)^{k-1}
T_k}{(2z)^{2k}}\\ 
\beta^{(n)}(z) &\sim& (-1)^n \left[\frac{n!}{2 z^{n+1}} + \sum_{k=1}^{\infty}
\frac{(-1)^{k-1}}{2^{2k}~~z^{2k+n}} T_k \frac{(2k+n-1)!}{(2k-1)!}\right]\\  
\beta(z) &\sim&
\frac{1}{2}\,\frac{1}{z}+\frac{1}{4}\,\frac{1}{z^2}-\frac{1}{8}\,\frac{1}{z^4}
+\frac{1}{4}\,\frac{1}{z^6}-{\frac 
{17}{16
}}\,\frac{1}{z^8}+{\frac {31}{4}}\,\frac{1}{z^{10}}-{\frac 
{691}{8}}\,\frac{1}{z^{12}}+{
\frac {5461}{4}}\,\frac{1}{z^{14}}
\nonumber\\
& & -{\frac 
{929569}{32}}\,\frac{1}{z^{16}}
+{\frac {
3202291}{4}}\,\frac{1}{z^{18}}-{\frac {221930581}{8}}\,\frac{1}{z^{20}}
+O\left(\frac{1}{z^{22}}\right)
\\
\beta'(z) &\sim&
-\frac{1}{2}\,\frac{1}{z^2}-\frac{1}{2}\,\frac{1}{z^3}
+\frac{1}{2}\,\frac{1}{z^5}-\frac{3}{2}\,\frac{1}{z^7}
+\frac{17}{2}\,\frac{1}{z^9}
-{\frac {155}{2}}\,\frac{1}{z^{11}}+{\frac 
{2073}{2}}\,\frac{1}{z^{13}}-{\frac {
38227}{2}}\,\frac{1}{z^{15}}
\nonumber\\ && 
+{\frac {929569}{2}}\,\frac{1}{z^{17}}-{\frac 
{28820619}{2}}\,\frac{1}{z^{19}}+{\frac {1109652905}{2}}\,\frac{1}{z^{21}}
+O\left(\frac{1}{z^{23}}\right)
\\
\beta^{(2)}(z) &\sim& 
\frac{1}{z^3}+\frac{3}{2}\,\frac{1}{z^4}-\frac{5}{2}\,\frac{1}{z^6}
+\frac{21}{2}\,\frac{1}{z^8}
-{\frac {153}{2}}\, \frac{1}{z^{10}}+{\frac 
{1705}{2}}\,\frac{1}{z^{12}}-{\frac {26949}{2}}\,\frac{1}{z^{14}}
+{\frac {573405}{2}}\,\frac{1}{z^{16}}
\nonumber\\ && 
-{\frac {15802673}{2}}\,\frac{1}{z^{18}}
+{\frac{547591761}{2}}\,\frac{1}{z^{20}}
+O\left(\frac{1}{z^{22}}\right)
\\
\beta^{(3)}(z) &\sim& 
-\frac{3}{z^4}-\frac{6}{z^5}+\frac{15}{z^7}
-\frac{84}{z^9}+\frac{765}{z^{11}}
-\frac{10230}{z^{13}}+\frac{188643}{z^{15}}
-\frac{4587240}{z^{17}}
\nonumber\\ && 
+\frac{142224057}{z^{19}}-\frac{5475917610}{z^{21}}
+O\left(\frac{1}{z^{23}}\right)
\\
\beta^{(4)}(z) &\sim& 
\frac{12}{z^{5}}+\frac{30}{z^{6}}-\frac{105}{z^{8}}+\frac{756}{z^{10}}
-\frac{8415}{z^{12}}
+\frac{132990}{z^{14}}-\frac{2829645}{z^{16}}+\frac{77983080}{z^{18}}
\nonumber\\ && 
-\frac{2702257083}{z^{20}}
+O\left(\frac{1}{z^{22}}\right)
\\
\beta^{(5)}(z) &\sim&   
-\frac{60}{z^{6}}-\frac{180}{z^{7}}+\frac{840}{z^{9}}
-\frac{7560}{z^{11}}+\frac{100980}{z^{13}}-\frac{1861860}{z^{15}}
+\frac{45274320}{z^{17}}-\frac{1403695440}{z^{19}}\nonumber\\ && 
+
\frac{54045141660}{z^{21}}
+O\left(\frac{1}{z^{23}}\right)
\\
F_1(z-1) &=& \Mvec\left[\frac{\ln(1+x)}{1+x}\right](z-1)\nonumber\\
&\sim&  
\Biggl[
\frac{1}{2} \frac{1}{z}
+\frac{1}{4} 
\frac{1}{z^2}
-\frac{1}{8}\,\frac{1}{z^4}
+\frac{1}{4}\,\frac{1}{z^6}
-{\frac {17}{16}}\,\frac{1}{z^8}
+{\frac {31}{4}}\,\frac{1}{z^{10}}
-{\frac {691}{8}}\,\frac{1}{z^{12}}
+{\frac {5461}{4}}\,\frac{1}{z^{14}}
\nonumber\\ 
& &
-{\frac {929569}{32}}\,\frac{1}{z^{16}}
+{\frac {3202291}{4}}\,\frac{1}{z^{18}}\Biggr] \ln(2)
\nonumber\\ &&
-\frac{1}{4}\,\frac{1}{z^2}
-\frac{1}{8}\,\frac{1}{z^3}
+\frac{3}{16}\,\frac{1}{z^4}
+\frac{3}{16}\,\frac{1}{z^5}
-{\frac {15}{32}}\,\frac{1}{z^6}
-\frac{5}{8}\,\frac{1}{z^7}
+{\frac {147}{64}}\,\frac{1}{z^8}
+{\frac {119}{32}}\,\frac{1}{z^9}
\nonumber\\ & &
-{\frac {1185}{64}}\,  \frac{1}{z^{10}}
-{\frac {279}{8}}\,    \frac{1}{z^{11}}
+{\frac {28479}{128}}\,\frac{1}{z^{12}}
+{\frac {7601}{16}}\,  \frac{1}{z^{13}}
-{\frac {238875}{64}}\,\frac{1}{z^{14}}
-{\frac {70993}{8}}\,  \frac{1}{z^{15}}
\nonumber\\ & &
+{\frac {21347283}{256}}\,   \frac{1}{z^{16}}
+{\frac {13943535}{64}}\,    \frac{1}{z^{17}}
-{\frac {306498525}{128}}\,  \frac{1}{z^{18}}
-{\frac {54438947}{8}}\,     \frac{1}{z^{19}}
+O\left(\frac{1}{z^{20}}\right)
\nonumber\\
\end{eqnarray}\begin{eqnarray}
F_2(z-1) &=& 
\Mvec\left\{\frac{\left[\ln^2(1+x)-\ln^2(2)\right]}{1-x}\right\}(z-1)
\nonumber\\
&\sim&  
\Biggl[
-\frac{1}{z}
-\frac{1}{4} \frac{1}{z^2}
+\frac{1}{12}\,\frac{1}{z^3}
+\frac{1}{16}\,\frac{1}{z^4}
-{\frac {11}{120}}\,\frac{1}{z^5}
-\frac{1}{16}\,\frac{1}{z^6}
+{\frac {19}{84}}\,\frac{1}{z^7}
+{\frac {9}{64}}\,\frac{1}{z^8}
-{\frac {247}{240}}\,\frac{1}{z^9}
\nonumber\\ & &
-{\frac {19}{32}}\,\frac{1}{z^{10}}
+{\frac {1013}{132}}\,\frac{1}{z^{11}}
+{\frac {269}{64}}\,\frac{1}{z^{12}}
-{\frac {940451}{10920}}\,\frac{1}{z^{13}}
-{\frac {1461}{32}}\,\frac{1}{z^{14}}
+{\frac {16369}{12}}\,\frac{1}{z^{15}}
\nonumber\\ & &
+{\frac {181561}{256}}\,\frac{1}{z^{16}}
-{\frac {236982223}{8160}}\,\frac{1}{z^{17}}
-{\frac {955231}{64}}\,\frac{1}{z^{18}}
+{\frac {1277626375}{1596}}\,\frac{1}{z^{19}}
\Biggr] \ln(2)
\nonumber\\ & & 
+\frac{1}{4}\,\frac{1}{z^2}-{\frac 
{5}{32}}\,\frac{1}{z^4}+\frac{1}{16}\,\frac{1}{z^5}+{\frac {13}{48}
}\,\frac{1}{z^6}-{\frac {9}{32}}\,\frac{1}{z^7}-{\frac 
{703}{768}}\,\frac{1}{z^8}+{
\frac {111}{64}}\,\frac{1}{z^9}+{\frac 
{3387}{640}}\,\frac{1}{z^{10}}
\nonumber\\ & &
-{\frac {995}
{64}}\,\frac{1}{z^{11}}-{\frac {18367}{384}}\,\frac{1}{z^{12}}+{\frac 
{25251}{128}}
\,\frac{1}{z^{13}}+{\frac {16932553}{26880}}\,\frac{1}{z^{14}}-{\frac 
{218421}{64}}
\,\frac{1}{z^{15}}
\nonumber\\ & &
-{\frac {23430373}{2048}}\,\frac{1}{z^{16}}+{\frac 
{19894795}{256}
}\,\frac{1}{z^{17}}+{\frac {2108066947}{7680}}\,\frac{1}{z^{18}}-{\frac 
{289304367}{128}}\,\frac{1}{z^{19}}
+O\left(\frac{1}{z^{20}}\right)
\nonumber\\
\\
F_3(z-1) &=& \Mvec\left[\frac{\ln^2(1+x)}{1+x}\right](z-1)\nonumber\\
&\sim&  
\Biggl[
\frac{1}{2}\,\frac{1}{z}+\frac{1}{4}\,\frac{1}{z^{2}}
-\frac{1}{8}\,\frac{1}{z^4}+\frac{1}{4}\,\frac{1}{z^6}
-{\frac {17}{16}}\,\frac{1}{z^8}
+{\frac {31}{4}}\,\frac{1}{z^{10}}
-{\frac {691}{8}}\,\frac{1}{z^{12}}
+{\frac {5461}{4}}\,\frac{1}{z^{14}}
\nonumber\\ & &
-{\frac {929569}{32}}\,\frac{1}{z^{16}}
+{\frac {3202291}{4}}\,\frac{1}{z^{18}}\Biggr] \ln^2(2)
\nonumber\\ & &
+ 
\Biggl[-\frac{1}{2}\,\frac{1}{z^2}-\frac{1}{4}\,\frac{1}{z^3}
+\frac{3}{8}\,\frac{1}{z^4}+\frac{3}{8}\,\frac{1}{z^5}
-{\frac {15}{16}}\,\frac{1}{z^6}-\frac{5}{4}\,\frac{1}{z^7}
+{\frac {147}{32}}\,\frac{1}{z^8}+{\frac {119}{
16}}\,\frac{1}{z^9}
\nonumber\\ & & 
-{\frac {1185}{32}}\,\frac{1}{z^{10}}-{\frac 
{279}{4}}\,\frac{1}{z^{11}}+{\frac {28479}{64}}\,\frac{1}{z^{12}}
+{\frac {7601}{8}}\,\frac{1}{z^{13}}-{
\frac {238875}{32}}\,\frac{1}{z^{14}}-{\frac 
{70993}{4}}\,\frac{1}{z^{15}}
\nonumber\\ 
& & 
+{\frac {21347283}{128}}\,\frac{1}{z^{16}}+{\frac 
{13943535}{32}}\,\frac{1}{z^{17}}-{\frac {
306498525}{64}}\,\frac{1}{z^{18}}-{\frac 
{54438947}{4}}\,\frac{1}{z^{19}}
\Biggr]\ln(2)
\nonumber\\ & &
+
\frac{1}{4}\,\frac{1}{z^3}-{\frac {9}{16}}\,\frac{1}{z^5}
+{\frac {5}{32}}\,\frac{1}{z^6}
+{\frac {75}{32}}\,\frac{1}{z^7}
-{\frac {49}{32}}\,\frac{1}{z^8}
-{\frac {1029}{64}}\,\frac{1}{z^9}
+{\frac {2259}{128}}\,\frac{1}{z^{10}}+{\frac 
{10665}{64}}\,\frac{1}{z^{11}}
\nonumber\\ && 
-{\frac {68057}{256}}\,\frac{1}{z^{12}}
-{\frac {313269}{128}}\,\frac{1}{z^{13}}
+{\frac {1338415}{256}}\,\frac{1}{z^{14}}
+{\frac {3105375}{64}}\,\frac{1}{z^{15}}
-{\frac {33703065}{256}}\,\frac{1}{z^{16}}
\nonumber\\ && 
-{\frac {320209245}{256}}\,\frac{1}{z^{17}}
+{\frac {2127156467}{512}}\,\frac{1}{z^{18}}
+{\frac {5210474925}{128}}\,\frac{1}{z^{19}}
+O\left(\frac{1}{z^{20}}\right)
\\
\hat{F}_4(z-1) &=& \Mvec\left[\frac{\Li_2(1-x)}{1-x}\right](z-1)\nonumber\\ 
&\sim&  
\frac{1}{z}
+\frac{1}{4 z^2}
-\frac{1}{36}\,\frac{1}{z^3}
-\frac{1}{24}\,\frac{1}{z^4}
+{\frac {7}{450}}\,\frac{1}{z^5}
+\frac{1}{40}\,\frac{1}{z^6}
-{\frac {38}{2205}}\,\frac{1}{z^7}
-{\frac {5}{168}}\,\frac{1}{z^{8}}+{\frac {11}{350}}\,\frac{1}{z^{9}}
\nonumber\\ & & 
+{\frac {7}{120}}\,\frac{1}{z^{10}}
-{\frac {3263}{38115}}\,\frac{1}{z^{11}}
-{\frac {15}{88}}\,\frac{1}{z^{12}}
+{\frac {13399637}{40990950}}\,\frac{1}{z^{13}}
+{\frac {7601}{10920}}\,\frac{1}{z^{14}}
-{\frac {8364}{5005}}\,\frac{1}{z^{15}}
\nonumber\\ & & 
-{\frac {91}{24}}\,\frac{1}{z^{16}}
+{\frac {1437423473}{130180050}}\,\frac{1}{z^{17}}
+{\frac {3617}{136}}\,\frac{1}{z^{18}}
-{\frac {177451280177}{1935088155}}\,\frac{1}{z^{19}}
+O\left(\frac{1}{z^{20}}\right)
\end{eqnarray}\begin{eqnarray}
\hat{F}_5(z-1) &=& \Mvec\left[\frac{\Li_2(1-x)}{1+x}\right](z-1)\nonumber\\ 
&\sim&  
\frac{1}{2}\,\frac{1}{z^2}+\frac{1}{4}\,\frac{1}{z^3}
-{\frac {7}{24}}\,\frac{1}{z^4}-\frac{1}{3}\,\frac{1}{z^5}
+{\frac {73}{120}}\,\frac{1}{z^6}+{\frac 
{31}{30}}\,\frac{1}{z^7}
-{\frac {439}{168}}\,\frac{1}{z^8}
-{\frac {206}{35}}\,\frac{1}{z^9}
\nonumber\\ & &
+{\frac {2293}{120}}\,\frac{1}{z^{10}}
+{\frac {11279}{210}}\,\frac{1}{z^{11}}
-{\frac {56255}{264}}\,\frac{1}{z^{12}}
-{\frac {829781}{1155}}\,\frac{1}{z^{13}}
+{\frac {36784003}{10920}}\,\frac{1}{z^{14}}
\nonumber\\ & &
+{\frac {132616441}{10010}}\,\frac{1}{z^{15}}
-{\frac {1720201}{24}}\,\frac{1}{z^{16}}
-{\frac {4837288492}{15015}}\,\frac{1}{z^{17}}
+{\frac {4029674381}{2040}}\,\frac{1}{z^{18}}
\nonumber\\ & &
+{\frac {5099210469253}{510510}}\,\frac{1}{z^{19}}
+O\left(\frac{1}{z^{20}}\right)
\\
\hat{F}_{6a}(z-1) &=& 
\Mvec\left[\frac{S_{1,2}(1-x)}{1-x}\right](z-1)\nonumber\\ 
&\sim&
 \frac{1}{4}\,     \frac{1}{z^2}
+\frac{1}{12}\,    \frac{1}{z^3}
-\frac{1}{16}\,    \frac{1}{z^4}
-\frac{1}{24}\,    \frac{1}{z^5}
+{\frac {7}{144}}\,\frac{1}{z^6}
+\frac{1}{24}\,    \frac{1}{z^7}
-{\frac {5}{72}}\, \frac{1}{z^8}
-{\frac {5}{72}}\, \frac{1}{z^9}
+{\frac {187}{1200}}\,\frac{1}{z^{10}}
\nonumber\\ & &
+{\frac {7}{40}}\,    \frac{1}{z^{11}}
-{\frac {91}{180}}\,  \frac{1}{z^{12}}
-\frac{5}{8}\,\frac{1}{z^{13}}
+{\frac {395903}{176400}}\,\frac{1}{z^{14}}
+{\frac {7601}{2520}}\,\frac{1}{z^{15}}
-{\frac {10999}{840}}\,\frac{1}{z^{16}}
-{\frac {455}{24}}\,   \frac{1}{z^{17}}
\nonumber\\ & &
+{\frac {2451931}{25200}}\,\frac{1}{z^{18}}
+{\frac {3617}{24}}\,\frac{1}{z^{19}}
+O\left(\frac{1}{z^{20}}\right)
\\
\hat{F}_{6b}(z-1) &=& 
\Mvec\left[\frac{S_{1,2}(1-x)}{1+x}\right](z-1)\nonumber\\ 
&\sim&  
\frac{1}{4}\,\frac{1}{z^3}
+\frac{1}{8}\,\frac{1}{z^4}-\frac{3}{8}\,\frac{1}{z^5}
-{\frac {5}{16}}\,\frac{1}{z^6}
+{\frac {29}{24}}\,\frac{1}{z^7}
+{\frac {133}{96}}\,\frac{1}{z^8}
-{\frac {167}{24}}\,\frac{1}{z^9}
-{\frac {163}{16}}\,\frac{1}{z^{10}}
+{\frac {2547}{40}}\,\frac{1}{z^{11}}
\nonumber\\  && 
+{\frac {9097}{80}}\,\frac{1}{z^{12}}
-{\frac {20455}{24}}\,\frac{1}{z^{13}}
-{\frac {431483}{240}}\,\frac{1}{z^{14}}
+{\frac {13204319}{840}}\,\frac{1}{z^{15}}
+{\frac {85692353}{2240}}\,\frac{1}{z^{16}}
\nonumber\\  && 
-{\frac {3058125}{8}}\,\frac{1}{z^{17}}
-{\frac {590412261}{560}}\,\frac{1}{z^{18}}
+{\frac {1422236953}{120}}\,\frac{1}{z^{19}}
+O\left(\frac{1}{z^{20}}\right)
\\
\hat{F}_7(z-1) &=& \Mvec\left[\frac{\Li_3(1-x)}{1-x}\right](z-1)\nonumber\\ 
&\sim&  
\frac{1}{z}+\frac{1}{8}\,\frac{1}{z^2}
-{\frac {11}{216}}\,\frac{1}{z^3}
-{\frac {1}{288}}\,\frac{1}{z^4}
+{\frac {1243}{54000}}\,\frac{1}{z^5}
-{\frac {49}{7200}}\,\frac{1}{z^6}
-{\frac {75613}{3704400}}\,\frac{1}{z^7}
+{\frac {599}{35280}}\,\frac{1}{z^8}
\nonumber\\ &&
+{\frac {234671}{7938000}}\,\frac{1}{z^9}
-{\frac {803}{16800}}\,\frac{1}{z^{10}}
-{\frac {4955857}{78262800}}\,\frac{1}{z^{11}}
+{\frac {53443}{304920}}\,\frac{1}{z^{12}}
+{\frac {921931911863}{4923832914000}}\,\frac{1}{z^{13}}
\nonumber\\ &&
-{\frac {449291}{535392}}\,\frac{1}{z^{14}}
-{\frac {23461249769}{32464832400}}\,\frac{1}{z^{15}}
+{\frac {1237447}{240240}}\,\frac{1}{z^{16}}
+{\frac {917870505450709}{265832869302000}}\,\frac{1}{z^{17}}
\nonumber\\ &&
-{\frac {82659252107}{2082880800}}\,\frac{1}{z^{18}}
-{\frac {959539811053709101}{50052680603125200}}\,\frac{1}{z^{19}}
+O\left(\frac{1}{z^{20}}\right)
\\
\hat{F}_8(z-1) &=& \Mvec\left[\frac{\Li_3(1-x)}{1+x}\right](z-1)\nonumber\\ 
&\sim&  
\frac{1}{2}\,\frac{1}{z^2}+\frac{1}{8}\,\frac{1}{z^3}
-{\frac {47}{144}}\,\frac{1}{z^4}
-{\frac {19}{144}}\,\frac{1}{z^5}
+{\frac {4931}{7200}}\,\frac{1}{z^6}
+{\frac {1339}{3600}}\,\frac{1}{z^7}
-{\frac {82769}{28224}}\,\frac{1}{z^8}
-{\frac {40357}{19600}}\,\frac{1}{z^9}
\nonumber\\ & &
+{\frac {119883}{5600}}\,\frac{1}{z^{10}}
+{\frac {1096889}{58800}}\,\frac{1}{z^{11}}
-{\frac {291002153}{1219680}}\,\frac{1}{z^{12}}
-{\frac {5317051741}{21344400}}\,\frac{1}{z^{13}}
\nonumber\\ & &
+{\frac {2473232407561}{655855200}}\,\frac{1}{z^{14}}
+{\frac {16564407196867}{3607203600}}\,\frac{1}{z^{15}}
-{\frac {51402420497}{640640}}\,\frac{1}{z^{16}}
\nonumber\\ & &
-{\frac {402766245627101}{3607203600}}\,\frac{1}{z^{17}}
+{\frac {4605756568109209}{2082880800}}\,\frac{1}{z^{18}}
+{\frac {400979739762637411}{115831315600}}\,\frac{1}{z^{19}}
\nonumber\\ & &
+O\left(\frac{1}{z^{20}}\right)
\end{eqnarray}\begin{eqnarray}
\hat{F}_{9}(z-1) &=& 
\Mvec\left[\frac{S_{1,3}(1-x)}{1-x}\right](z-1)\nonumber\\ 
&\sim&   
 \frac{1}{9}\,\frac{1}{z^3}
+\frac{1}{24}\,\frac{1}{z^4}
-{\frac {13}{180}}\,\frac{1}{z^5}
-\frac{1}{24}\,\frac{1}{z^6}
+{\frac {11}{126}}\,\frac{1}{z^7}
+\frac{1}{16}\,\frac{1}{z^8}
-{\frac {68}{405}}\,\frac{1}{z^9}
-{\frac {5}{36}}\,\frac{1}{z^{10}}
+{\frac {281}{594}}\,\frac{1}{z^{11}}
\nonumber\\ & &
+{\frac {7}{16}}\,\frac{1}{z^{12}}
-{\frac {22696}{12285}}\,\frac{1}{z^{13}}
-{\frac {15}{8}}\,\frac{1}{z^{14}}
+{\frac {38857}{4050}}\,\frac{1}{z^{15}}
+{\frac {7601}{720}}\,\frac{1}{z^{16}}
-{\frac {147176}{2295}}\,\frac{1}{z^{17}}
\nonumber\\ & &
-{\frac {455}{6}}\,\frac{1}{z^{18}}
+{\frac {96472231}{179550}}\,\frac{1}{z^{19}}
+O\left(\frac{1}{z^{20}}\right)
\\
\hat{F}_{10}(z-1) &=& 
\Mvec\left[\frac{S_{1,3}(1-x)}{1+x}\right](z-1)\nonumber\\ 
&\sim&   
\frac{1}{6}\,\frac{1}{z^4}
+\frac{1}{12}\,\frac{1}{z^5}
-{\frac {11}{24}}\,\frac{1}{z^6}
-\frac{1}{3}\,\frac{1}{z^7}
+{\frac {101}{48}}\,\frac{1}{z^8}
+\frac{2}{z^9}-{\frac {563}{36}}\,\frac{1}{z^{10}}
-{\frac {166}{9}}\,\frac{1}{z^{11}}
+{\frac {2801}{16}}\,\frac{1}{z^{12}}
\nonumber\\ & &
+{\frac {2225}{9}}\,\frac{1}{z^{13}}
-{\frac {66475}{24}}\,\frac{1}{z^{14}}
-{\frac {41047}{9}}\,\frac{1}{z^{15}}
+{\frac {42441911}{720}}\,\frac{1}{z^{16}}
+{\frac {4991108}{45}}\,\frac{1}{z^{17}}
\nonumber\\ & &
-{\frac {9747745}{6}}\,\frac{1}{z^{18}}
-{\frac {154748044}{45}}\,\frac{1}{z^{19}}
+O\left(\frac{1}{z^{20}}\right)
\\
\hat{F}_{11}(z-1) &=& 
\Mvec\left[\frac{S_{2,2}(1-x)}{1-x}\right](z-1)\nonumber\\ 
&\sim&  
+\frac{1}{8}\,\frac{1}{z^2}
-{\frac {1}{72}}\,\frac{1}{z^3}
-{\frac {7}{192}}\,\frac{1}{z^4}
+{\frac {31}{1440}}\,\frac{1}{z^5}
+{\frac {41}{1728}}\,\frac{1}{z^6}
-{\frac {23}{672}}\,\frac{1}{z^7}
-{\frac {85}{3456}}\,\frac{1}{z^8}
\nonumber\\ & &
+{\frac {487}{6480}}\,\frac{1}{z^9}
+{\frac {103}{3000}}\,\frac{1}{z^{10}}
-{\frac {54217}{237600}}\,\frac{1}{z^{11}}
-{\frac {4283}{86400}}\,\frac{1}{z^{12}}
+{\frac {18341}{19656}}\,\frac{1}{z^{13}}
\nonumber\\ & &
-{\frac {1730749}{74088000}}\,\frac{1}{z^{14}}
-{\frac {15856763}{3175200}}\,\frac{1}{z^{15}}
+{\frac {1958773}{1411200}}\,\frac{1}{z^{16}}
+{\frac {8758993}{257040}}\,\frac{1}{z^{17}}
\nonumber\\ & &
-{\frac {98100199}{5292000}}\,\frac{1}{z^{18}}
-{\frac {118909349}{410400}}\,\frac{1}{z^{19}}
+O\left(\frac{1}{z^{20}}\right)
\\
\hat{F}_{12}(z-1) &=& 
\Mvec\left[\frac{S_{2,2}(1-x)}{1+x}\right](z-1)\nonumber\\ 
&\sim&   
+\frac{1}{8}\,\frac{1}{z^3}
-\frac{1}{48}\,\frac{1}{z^4}
-{\frac {19}{96}}\,\frac{1}{z^5}
+{\frac {17}{192}}\,\frac{1}{z^6}
+{\frac {181}{288}}\,\frac{1}{z^7}
-{\frac {523}{1152}}\,\frac{1}{z^8}
-{\frac {515}{144}}\,\frac{1}{z^9}
+{\frac {1001}{288}}\,\frac{1}{z^{10}}
\nonumber\\ 
& &
+{\frac {234511}{7200}}\,\frac{1}{z^{11}}
-{\frac {376301}{9600}}\,\frac{1}{z^{12}}
-{\frac {15671}{36}}\,\frac{1}{z^{13}}
+{\frac {8946367}{14400}}\,\frac{1}{z^{14}}
+{\frac {314590397}{39200}}\,\frac{1}{z^{15}}
\nonumber\\ 
& &
-{\frac {4666625479}{352800}}\,\frac{1}{z^{16}}
-{\frac {196698251}{1008}}\,   \frac{1}{z^{17}}
+{\frac {21437856917}{58800}}\,\frac{1}{z^{18}}
+{\frac {101639524687}{16800}}\,\frac{1}{z^{19}}
\nonumber\\ & &
+O\left(\frac{1}{z^{20}}\right)
\\
\hat{F}_{13}(z-1) &=& 
\Mvec\left[\frac{\Li_2^2(1-x)}{1-x}\right](z-1)\nonumber\\ 
&\sim&   
\frac{1}{z^2}
-{\frac {7}{24}}\,\frac{1}{z^4}
+\frac{1}{12}\,\frac{1}{z^5}
+{\frac {223}{1080}}\,\frac{1}{z^6}
-{\frac {7}{45}}\,\frac{1}{z^7}
-{\frac {3767}{15120}}\,\frac{1}{z^8}
+{\frac {38}{105}}\,\frac{1}{z^9}
+{\frac {14327}{31500}}\,\frac{1}{z^{10}}
\nonumber\\ &&
-{\frac {198}{175}}\,\frac{1}{z^{11}}
-{\frac {138673}{118800}}\,\frac{1}{z^{12}}
+{\frac {3263}{693}}\,\frac{1}{z^{13}}
+{\frac {5265804043}{1324323000}}\,\frac{1}{z^{14}}
-{\frac{13399637}{525525}}\,\frac{1}{z^{15}}
\nonumber\\ &&
-{\frac {143341487}{8408400}}\,\frac{1}{z^{16}}
+{\frac {25092}{143}}\,\frac{1}{z^{17}}
+{\frac {34809672614}{402026625}}\,\frac{1}{z^{18}}
-{\frac {5749693892}{3828825}}\,\frac{1}{z^{19}}
+O\left(\frac{1}{z^{20}}\right)
\nonumber\\ 
\end{eqnarray}\begin{eqnarray}
\hat{F}_{14}(z-1) &=& 
\Mvec\left[\frac{\Li_2^2(1-x)}{1+x}\right](z-1)\nonumber\\ 
&\sim&   
\frac{1}{z^3}
-{\frac {19}{12}}\,\frac{1}{z^5}
+{\frac {5}{24}}\,\frac{1}{z^6}
+{\frac {457}{90}}\,\frac{1}{z^7}
-{\frac {917}{720}}\,\frac{1}{z^8}
-{\frac {9152}{315}}\,\frac{1}{z^9}
+{\frac {4261}{420}}\,\frac{1}{z^{10}}
+{\frac {13257}{50}}\,\frac{1}{z^{11}}
\nonumber\\ &&
-{\frac {970387}{8400}}\,\frac{1}{z^{12}}
-{\frac {12286094}{3465}}\,\frac{1}{z^{13}}
+{\frac {254353619}{138600}}\,\frac{1}{z^{14}}
+{\frac {206155893467}{3153150}}\,\frac{1}{z^{15}}
\nonumber\\ &&
-{\frac {109568208507}{2802800}}\,\frac{1}{z^{16}}
-{\frac {1591422384}{1001}}\,\frac{1}{z^{17}}
+{\frac {377551202267}{350350}}\,\frac{1}{z^{18}}
\nonumber\\ &&
+{\frac {377455058454233}{7657650}}\,\frac{1}{z^{19}}
+O\left(\frac{1}{z^{20}}\right)
\\
\hat{F}_{15}(z-1) &=& \Mvec\left[\frac{\ln(x) S_{1,2}(-x) - 
\Li_2^2(-x)/2+\zeta_2^2/8}{1-x}\right](z-1) \nonumber\\ &=&
\left(-\frac{1}{8}\,\zeta_3+\frac{1}{2}\,\zeta_2\,\ln(2)\right) \frac{1}{z}
+\left(-\frac{1}{16}\,\zeta_3-\frac{1}{8}\,\zeta_2+\frac{1}{4}\,
  \zeta_2\,\ln(2)\right)\frac{1}{z^{2}}
\nonumber\\ &&
+\left(-\frac{1}{48}\,\zeta_3-\frac{1}{12}\,\zeta_2+\frac{1}{12}\,\zeta_2\,
  \ln(2)\right)\frac{1}{z^{3}}
+\frac{1}{16 z^4}
+ \left( {\frac {1}{240}}\,\zeta_3+{\frac {7}{240}}\,\zeta_2-{\frac 
{1}{60}}\,\zeta_2\,\ln(2) \right) 
\frac{1}{z^5}
\nonumber\\ &&
+ \left( -\frac{1}{8}-{\frac {1}{96}}
\,\zeta_2 \right) 
\frac{1}{z^6}
+ \left( \frac{1}{16}-{\frac {1}{336}}\,\zeta_3-{\frac {23}{672}}\,
\zeta_2+{\frac {1}{84}}\,\zeta_2\,\ln(2) \right) \frac{1}{z^7}
\nonumber\\ &&
+ \left( \frac{1}{24}\,\zeta_2+{
\frac {13}{32}} \right) \frac{1}{z^8}
+ \left( -\frac{1}{2}-{\frac {1}{60}}\,\zeta_2\,\ln(2)+{\frac {
13}{160}}\,\zeta_2+{\frac {1}{240}}\,\zeta_3 \right)\frac{1}{z^9}
\nonumber\\
&&
+ \left( -{\frac {73}{320}}
\,\zeta_2-{\frac {33}{16}} \right) \frac{1}{z^{10}}
+ \left( {\frac {5}{132}}\,\zeta_2\,\ln(2)
-{\frac {719}{2112}}\,\zeta_2-{\frac {5}{528}}\,\zeta_3+{\frac {73}{16}} \right) 
\frac{1}{z^{11}}
\nonumber\\ &&
+ \left( {\frac {11}{6}}\,\zeta_2+{\frac {1517}{96}} \right) \frac{1}{z^{12}}
+ \left( -{\frac {
691}{5460}}\,\zeta_2\,\ln(2)+{\frac {203923}{87360}}\,\zeta_2
+{\frac {691}{21840}}\,\zeta_3-{\frac {109}{2}} \right)\frac{1}{z^{13}} 
\nonumber\\ &&
+ \left( -{\frac {71173}{3360}}\,\zeta_2-{\frac 
{527}{3}} \right) \frac{1}{z^{14}}
+ \left( {\frac {7}{12}}\,\zeta_2\,\ln(2)-{\frac {1571}{64}}
\,\zeta_2-{\frac {7}{48}}\,\zeta_3+{\frac {13821}{16}} \right) 
\frac{1}{z^{15}}
\nonumber\\ &&
+ \left( {\frac {
2713}{8}}\,\zeta_2+{\frac {261131}{96}} \right) \frac{1}{z^{16}}
+ \left( -17899-{\frac {3617}{
1020}}\,\zeta_2\,\ln(2)+{\frac {6074951}{16320}}\,
\zeta_2+{\frac {3617}{4080}}\,\zeta_3 \right) \frac{1}{z^{17}}
\nonumber\\ &&
+ \left( -{\frac {13914599}{1920}}\,\zeta_2-{\frac {2709997}{48}}
 \right) \frac{1}{z^{18}}
\nonumber\\ &&
+ \left( {\frac {43867}{1596}}\,\zeta_2\,\ln(2)-{\frac {393894905}{
51072}}\,\zeta_2-{\frac {43867}{6384}}\,\zeta_3+{\frac {7590505}{16}} \right) 
\frac{1}{z^{19}}
+O\left(\frac{1}{z^{20}}\right)
\\
\hat{F}_{16}(z-1) &=& \Mvec\left[\frac{\ln(x) S_{1,2}(-x) - 
\Li_2^2(-x)/2}{1+x}\right](z-1) \nonumber\\ &=& 
-{\frac {5}{32}}\,{\frac {\zeta_4}{z}}
+ \left( -\frac{1}{16}\,\zeta_3+\frac{1}{4}\,\zeta_2\,\ln(2)-{\frac 
{5}{64}}\,\zeta_4 \right) 
\frac{1}{z^2}
+\left(
-\frac{1}{16}\,\zeta_3+\frac{1}{4}\,\zeta_2\,\ln(2)-\frac{1}{8}\,\zeta_2\right) 
\frac{1}{z^{3}}
\nonumber\\ &&
+ \left( {\frac {5}{128}}\,\zeta_4-\frac{1}{8}\,\zeta_2 \right) 
\frac{1}{z^4}
+\left(\frac{1}{16}\,\zeta_3-\frac{1}{4}\,\zeta_2\,\ln(2)+
\frac{1}{8}+\frac{1}{8}\,\zeta_2\right) \frac{1}{z^{5}}
+ \left( {\frac {9}{32}}\,\zeta_2-{\frac {5}{64}}\,\zeta_4 \right) \frac{1}{z^6}
\nonumber\\ &&
+ \left( \frac{3}{4}\,\zeta_2\,\ln(2)-{\frac 
{13}{32}}\,\zeta_2-\frac{3}{16}\,\zeta_3-{\frac {11}{16}}
 \right) \frac{1}{z^7}
+ \left( -\frac{5}{4}\,\zeta_2+{\frac {85}{256}}\,\zeta_4+{\frac {7}{32}} 
\right) 
\frac{1}{z^8}
\nonumber\\ &&
+ \left( -{\frac {17}{4}}\,\zeta_2\,\ln(2)+{\frac {39}{16}}\,\zeta_2
+{\frac {17}{16}}\,\zeta_3+{\frac {41}{8}} \right) \frac{1}{z^9}
+ \left( {\frac {595}{64}}\,\zeta_2-{\frac {155}{64}}\,\zeta_4-{\frac 
{57}{16}} \right) 
\frac{1}{z^{10}}
\nonumber\\ &&
+ \left( {\frac {155}{4}}\,\zeta_2\,\ln(2)-{\frac {1477}{64}}\,\zeta_2-{\frac 
{155}{16}}\,\zeta_3-{
\frac {867}{16}} \right) 
\frac{1}{z^{11}}
\nonumber\\ &&
+ \left( -{\frac {837}{8}}\,\zeta_2+{
\frac {1925}{32}}+{\frac {3455}{128}}\,\zeta_4 \right) 
\frac{1}{z^{12}}
\nonumber
\end{eqnarray}\begin{eqnarray} &&
+ \left( -{\frac {2073}{4}}\,\zeta_2\,\ln(2)+{\frac {20269}{64}}\,\zeta_2+{
\frac {2073}{16}}\,\zeta_3+{\frac {6341}{8}} \right)
\frac{1}{z^{13}}
\nonumber\\ &&
+ \left( {\frac {53207}{32}}\,\zeta_2-{\frac {27305}{64}}\,\zeta_4-{\frac {9841}{8
}} \right) 
\frac{1}{z^{14}}
\nonumber\\ &&
+ \left( {\frac {38227}{4}}\,\zeta_2\,\ln(2)-{
\frac {380733}{64}}\,\zeta_2-{\frac {38227}{16}}\,\zeta_3-{\frac {248063
}{16}} \right) 
\frac{1}{z^{15}}
\nonumber\\ &&
+ \left( -{\frac {70993}{2}}\,\zeta_2+{\frac {
4647845}{512}}\,\zeta_4+{\frac {998859}{32}} \right) \frac{1}{z^{16}}
\nonumber\\ &&
+ \left( -{
\frac {929569}{4}}\,\zeta_2\,\ln(2)+{\frac {4693531}{32}}\,\zeta_2+{
\frac {929569}{16}}\,\zeta_3+{\frac {3143445}{8}} \right) \frac{1}{z^{17}}
\nonumber\\ &&
+ \left( {\frac {125491815}{128}}\,\zeta_2-{\frac {16011455}{64}}\,\zeta_4-{\frac 
{15651067}{16}} \right) \frac{1}{z^{18}}
\nonumber\\ &&
+ \left( {\frac {28820619}{4}}\,
\zeta_2\,\ln(2)-{\frac {588338053}{128}}\,\zeta_2-{\frac {28820619}{
16}}\,\zeta_3-{\frac {200948819}{16}} \right) \frac{1}{z^{19}}
\nonumber\\ && + O\left(\frac{1}{z^{20}}\right)~,
\\
\hat{F}_{17}(z-1) &=& 
\Mvec\left[\frac{\Li_4(1-x)}{1-x}\right](z-1)\nonumber\\ 
&\sim&   
 \frac{1}{z} +\frac{1}{16}\,\frac{1}{z^2} 
-{\frac {49}{1296}}\,\frac{1}{z^3} 
+{\frac {41}{3456}}\,\frac{1}{z^4} 
+{\frac {26291}{3240000}}\,\frac{1}{z^5} 
-{\frac {1921}{144000}}\,\frac{1}{z^6} 
+{\frac {845233}{1555848000}}\,\frac{1}{z^7} 
\nonumber\\ &&
+{\frac {1048349}{59270400}}\,\frac{1}{z^8} 
-{\frac {60517579}{5000940000}}\,\frac{1}{z^9} 
-{\frac {50233}{1587600}}\,\frac{1}{z^{10}} 
+{\frac {506605371959}{9762501672000}}\,\frac{1}{z^{11}} 
\nonumber\\ &&
+{\frac {823605863}{11269843200}}\,\frac{1}{z^{12}} 
-{\frac {53797712101337483}{221794053611130000}}\,\frac{1}{z^{13}} 
-{\frac {7784082036337}{39390663312000}}\,\frac{1}{z^{14}} 
\nonumber\\ &&
+{\frac {8049010408144441}{5849513501832000}}\,\frac{1}{z^{15}} 
+{\frac {246319059461}{519437318400}}\,\frac{1}{z^{16}} 
-{\frac {3910018782537447618421}{407131014322092060000}}\,\frac{1}{z^{17}}
\nonumber\\ &&
+{\frac {1090400590625849}{1063331477208000}}\,\frac{1}{z^{18}}
+{\frac {715799726422332035922796571}{8738918739347894481384000}}\,
        \frac{1}{z^{19}} 
+O\left(\frac{1}{z^{20}}\right) \nonumber\\
\\
\hat{F}_{18}(z-1) &=& 
\Mvec\left[\frac{\Li_4(1-x)}{1+x}\right](z-1)\nonumber\\ 
&\sim&   
 \frac{1}{2}\,\frac{1}{z^2}
+\frac{1}{16}\,\frac{1}{z^3}
-{\frac {265}{864}}\,\frac{1}{z^4}
-{\frac {67}{1728}}\,\frac{1}{z^5}
+{\frac {265597}{432000}}\,\frac{1}{z^6}
+{\frac {19043}{216000}}\,\frac{1}{z^7}
-{\frac {30715003}{11854080}}\,\frac{1}{z^8}
\nonumber\\ &&
-{\frac {5838257}{12348000}}\,\frac{1}{z^9}
+{\frac {598134997}{31752000}}\,\frac{1}{z^{10}}
+{\frac {954094949}{222264000}}\,\frac{1}{z^{11}}
-{\frac {262665087967}{1252204800}}\,\frac{1}{z^{12}}
\nonumber\\ &&
-{\frac {2125323437719}{36979173000}}\,\frac{1}{z^{13}}
+{\frac {130573212174747827}{39390663312000}}\,\frac{1}{z^{14}}
+{\frac {689293512543172297}{649945944648000}}\,\frac{1}{z^{15}}
\nonumber\\ &&
-{\frac {9158702639186257}{129859329600}}\, \frac{1}{z^{16}}
-{\frac {8382716878733905193}{324972972324000}}\,\frac{1}{z^{17}}
\nonumber\\ &&
+{\frac {1033390862857917785329}{531665738604000}}\,\frac{1}{z^{18}}
+{\frac {2553947932811624174969383}{3193184426055624000}}\,\frac{1}{z^{19}}
+O\left(\frac{1}{z^{20}}\right)
\nonumber\\
\end{eqnarray}

The asymptotic representations of the subsidiary functions $H_{1,(2,3)}(z)$ are given 
by~:
\begin{eqnarray}
{H}_{1}(z) &=& 
\Mvec\left[\frac{2 S_{1,2}(-x) + \ln(1+x) \Li_2(-x)}{1+x}\right](z)\nonumber 
\end{eqnarray}\begin{eqnarray}
&\sim& \Biggl[
 {\frac {1}{8}}\frac{1}{z^{4}}
-{\frac {1}{2}}\frac{1}{z^{6}}
+{\frac {3}{32}}\frac{1}{z^{7}}
+{\frac {191}{64}}\frac{1}{z^{8}}
-{\frac {5}{4}}\frac{1}{z^{9}}
-{\frac {855}{32}}\frac{1}{z^{10}}
+{\frac {1157}{64}}\frac{1}{z^{11}}
+{\frac {43829}{128}}\frac{1}{z^{12}}
\nonumber
\\ 
&&
-{\frac {10433}{32}}\frac{1}{z^{13}}
-{\frac {768023}{128}}\frac{1}{z^{14}}
+{\frac {238221}{32}}\frac{1}{z^{15}}
+{\frac {17710671}{128}}\frac{1}{z^{16}}
-{\frac {3407843}{16}}\frac{1}{z^{17}}
\nonumber
\\ 
&&
-{\frac {260505329}{64}}\frac{1}{z^{18}}
{\frac {961880921}{128}}\frac{1}{z^{19}}
\Biggr]
\nonumber\\ &&
+ \Biggl[
-{\frac {1}{4}}\frac{1}{z^{3}} 
-{\frac {1}{8}}\frac{1}{z^{4}}
+{\frac {1}{2}}\frac{1}{z^{5}}
+{\frac {3}{8}}\frac{1}{z^{6}}
-2\,\frac{1}{z^{7}}
-{\frac {15}{8}}\frac{1}{z^{8}}
+{\frac {27}{2}}\frac{1}{z^{9}}
+{\frac {119}{8}}\frac{1}{z^{10}}
\nonumber\\ &&
-139\,\frac{1}{z^{11}}
-{\frac {1395}{8}}\frac{1}{z^{12}}
+{\frac{4073}{2}}\frac{1}{z^{13}}
+{\frac {22803}{8}}\frac{1}{z^{14}}
-40356\,\frac{1}{z^{15}}
-{\frac {496951}{8}}\frac{1}{z^{16}}
\nonumber\\ &&
+{\frac {2081719}{2}}\frac{1}{z^{17}}
+{\frac {13943535}{8}}\frac{1}{z^{18}}
-33908573\,\frac{1}{z^{19}}
\Biggr] \ln(2) 
\nonumber\\ &&
+ \Biggl[
{\frac {1}{8}}\frac{1}{z^{2}} 
+{\frac {1}{16}}\frac{1}{z^{3}}
-{\frac {3}{32}}\frac{1}{z^{4}}
-{\frac {3}{32}}\frac{1}{z^{5}}
+{\frac {15}{64}}\frac{1}{z^{6}}
+{\frac {5}{16}}\frac{1}{z^{7}}
-{\frac {147}{128}}\frac{1}{z^{8}}
-{\frac {119}{64}}\frac{1}{z^{9}}
\nonumber\\ &&
+{\frac {1185}{128}}\frac{1}{z^{10}}
+{\frac {279}{16}}\frac{1}{z^{11}}
-{\frac {28479}{256}}\frac{1}{z^{12}}
-{\frac {7601}{32}}\frac{1}{z^{13}}
+{\frac {238875}{128}}\frac{1}{z^{14}}
+{\frac {70993}{16}}\frac{1}{z^{15}}
\nonumber\\ &&
-{\frac {21347283}{512}}\frac{1}{z^{16}}
-{\frac {13943535}{128}}\frac{1}{z^{17}}
+{\frac {306498525}{256}}\frac{1}{z^{18}}
+{\frac {54438947}{16}}\frac{1}{z^{19}}
\Biggr] \zeta_2
\nonumber\\ &&
+ \Biggl[
{\frac {1}{8}}\frac{1}{z}
+{\frac {1}{16}}\frac{1}{z^{2}}
-{\frac {1}{32}}\frac{1}{z^{4}}
+{\frac {1}{16}}\frac{1}{z^{6}}
-{\frac {17}{64}}\frac{1}{z^{8}}
+{\frac {31}{16}}\frac{1}{z^{10}}
-{\frac {691}{32}}\frac{1}{z^{12}}
+{\frac {5461}{16}}\frac{1}{z^{14}}
\nonumber\\ &&
-{\frac {929569}{128}}\frac{1}{z^{16}}
+{\frac {3202291}{16}}\frac{1}{z^{18}}
\Biggr] \zeta_3 
\nonumber\\ &&
+
\Biggl[
-{\frac {1}{4}}\frac{1}{z}
-{\frac {1}{8}}\frac{1}{z^{2}}
+{\frac {1}{16}}\frac{1}{z^{4}}
-{\frac {1}{8}}\frac{1}{z^{6}}
+{\frac {17}{32}}\frac{1}{z^{8}}
-{\frac {31}{8}}\frac{1}{z^{10}}
+{\frac {691}{16}}\frac{1}{z^{12}}
-{\frac {5461}{8}}\frac{1}{z^{14}}
\nonumber\\ &&
+{\frac {929569}{64}}\frac{1}{z^{16}}
-{\frac {3202291}{8}}\frac{1}{z^{18}}
\Biggr] \zeta_2 \ln(2) + O\left(\frac{1}{z^{20}}\right)
\end{eqnarray}\begin{eqnarray}
{H}_{2}(z) &=& 
\Mvec\left[\frac{\ln(1+x) \Li_3(-x) + \Li^2_2(-x)}{1+x}\right](z)\nonumber\\ 
&\sim& \Biggl[
-{\frac {3}{8}}\frac{1}{z} 
-{\frac {3}{16}}\frac{1}{z^{2}}
+{\frac {3}{32}}\frac{1}{z^{4}}
-{\frac {3}{16}}\frac{1}{z^{6}}
+{\frac {51}{64}}\frac{1}{z^{8}}
-{\frac {93}{16}}\frac{1}{z^{10}}
+{\frac {2073}{32}}\frac{1}{z^{12}}
-{\frac {16383}{16}}\frac{1}{z^{14}}
\nonumber\\ &&
+{\frac {2788707}{128}}\frac{1}{z^{16}}
-{\frac {9606873}{16}}\frac{1}{z^{18}}
\Biggr] \ln(2) \zeta_3
\nonumber
\\ 
&&
+ \Biggr[
 {\frac {1}{16}}\frac{1}{z} 
+{\frac {1}{32}}\frac{1}{z^{2}}
-{\frac {1}{64}}\frac{1}{z^{4}}
+{\frac {1}{32}}\frac{1}{z^{6}}
-{\frac {17}{128}}\frac{1}{z^{8}}
+{\frac {31}{32}}\frac{1}{z^{10}}
-{\frac {691}{64}}\frac{1}{z^{12}}
+{\frac {5461}{32}}\frac{1}{z^{14}}
\nonumber
\\ 
&&
-{\frac {929569}{256}}\frac{1}{z^{16}}
+{\frac{3202291}{32}}\frac{1}{z^{18}}
\Biggr] \zeta_2^2
\nonumber
\end{eqnarray}\begin{eqnarray}
&&
+ \Biggl[
-{\frac {1}{8}}\frac{1}{z^{3}}
-{\frac {1}{16}}\frac{1}{z^{4}}
+{\frac {1}{4}}\frac{1}{z^{5}}
+{\frac {3}{16}}\frac{1}{z^{6}}
-\frac{1}{z^{7}}
-{\frac {15}{16}}\frac{1}{z^{8}}
+{\frac {27}{4}}\frac{1}{z^{9}}
+{\frac {119}{16}}\frac{1}{z^{10}}
-{\frac {139}{2}}\frac{1}{z^{11}}
\nonumber
\end{eqnarray}\begin{eqnarray}
&&
-{\frac {1395}{16}}\frac{1}{z^{12}}
+{\frac {4073}{4}}\frac{1}{z^{13}}
+{\frac {22803}{16}}\frac{1}{z^{14}}
-20178\,\frac{1}{z^{15}}
-{\frac {496951}{16}}\frac{1}{z^{16}}
+{\frac {2081719}{4}}\frac{1}{z^{17}}
\nonumber\\ &&
+{\frac {13943535}{16}}\frac{1}{z^{18}}
-{\frac {33908573}{2}}\frac{1}{z^{19}}
\Biggr] \zeta_2
\nonumber\\ &&
+ \Biggl[
 {\frac {3}{16}}\frac{1}{z^{2}}
+{\frac {3}{32}}\frac{1}{z^{3}}
-{\frac {9}{64}}\frac{1}{z^{4}}
-{\frac {9}{64}}\frac{1}{z^{5}}
+{\frac {45}{128}}\frac{1}{z^{6}}
+{\frac {15}{32}}\frac{1}{z^{7}}
-{\frac {441}{256}}\frac{1}{z^{8}}
-{\frac {357}{128}}\frac{1}{z^{9}}
\nonumber\\ &&
+{\frac {3555}{256}}\frac{1}{z^{10}}
+{\frac {837}{32}}\frac{1}{z^{11}}
-{\frac {85437}{512}}\frac{1}{z^{12}}
-{\frac {22803}{64}}\frac{1}{z^{13}}
+{\frac {716625}{256}}\frac{1}{z^{14}}
+{\frac {212979}{32}}\frac{1}{z^{15}}
\nonumber\\ &&
-{\frac {64041849}{1024}}\frac{1}{z^{16}}
-{\frac {41830605}{256}}\frac{1}{z^{17}}
+{\frac {919495575}{512}}\frac{1}{z^{18}}
+{\frac {163316841}{32}}\frac{1}{z^{19}}
\Biggr] \zeta_3
\nonumber\\ &&
+ \Biggl[
 {\frac {1}{4}}\frac{1}{z^{4}}
+{\frac {1}{8}}\frac{1}{z^{5}}
-{\frac {15}{16}}\frac{1}{z^{6}}
-{\frac {5}{8}}\frac{1}{z^{7}}
+{\frac {175}{32}}\frac{1}{z^{8}}
+{\frac {35}{8}}\frac{1}{z^{9}}
-{\frac {777}{16}}\frac{1}{z^{10}}
-{\frac {357}{8}}\frac{1}{z^{11}}
\nonumber\\ &&
+{\frac {9933}{16}}\frac{1}{z^{12}}
+{\frac {5115}{8}}\frac{1}{z^{13}}
-{\frac {174031}{16}}\frac{1}{z^{14}}
-{\frac {98813}{8}}{z^{15}}
+{\frac {16066011}{64}}\frac{1}{z^{16}}
+{\frac {2484755}{8}}\frac{1}{z^{17}}
\nonumber\\ &&
-{\frac {118304037}{16}}\frac{1}{z^{18}}
-{\frac {79013365}{8}}\frac{1}{z^{19}}
\Biggr] \ln(2) + O\left(\frac{1}{z^{20}}\right)
\\
{H}_{3}(z) &=& 
\Mvec\left[\frac{\ln^3(1+x)}{1+x}\right](z)\nonumber\\ 
&\sim& \Biggl[
-{\frac {3}{8}}\frac{1}{z^{4}}
+{\frac {3}{8}}\frac{1}{z^{5}}
+{\frac {45}{32}}\frac{1}{z^{6}}
-{\frac {195}{64}}\frac{1}{z^{7}}
-{\frac {945}{128}}\frac{1}{z^{8}}
+{\frac {1869}{64}}\frac{1}{z^{9}}
+{\frac {3591}{64}}\frac{1}{z^{10}}
-{\frac {95265}{256}}\frac{1}{z^{11}}
\nonumber\\ &&
-{\frac {303435}{512}}\frac{1}{z^{12}}
+{\frac {3227961}{512}}\frac{1}{z^{13}}
+{\frac {8474895}{1024}}\frac{1}{z^{14}}
-{\frac {71170281}{512}}\frac{1}{z^{15}}
-{\frac {297211005}{2048}}\frac{1}{z^{16}}
\nonumber\\ &&
+{\frac {1993710885}{512}}\frac{1}{z^{17}}
+{\frac {3079871487}{1024}}\frac{1}{z^{18}}
-{\frac {138826111977}{1024}}\frac{1}{z^{19}}
\Biggr]
\nonumber\\ &&
+\Biggl[
{\frac {3}{4}}\frac{1}{z^{3}} 
-{\frac {27}{16}}\frac{1}{z^{5}}
+{\frac {15}{32}}\frac{1}{z^{6}}
+{\frac {225}{32}}\frac{1}{z^{7}}
-{\frac {147}{32}}\frac{1}{z^{8}}
-{\frac {3087}{64}}\frac{1}{z^{9}}
+{\frac {6777}{128}}\frac{1}{z^{10}}
+{\frac {31995}{64}}\frac{1}{z^{11}}
\nonumber\\ &&
-{\frac {204171}{256}}\frac{1}{z^{12}}
-{\frac {939807}{128}}\frac{1}{z^{13}}
+{\frac {4015245}{256}}\frac{1}{z}^{14}
+{\frac {9316125}{64}}\frac{1}{z^{15}}
-{\frac {101109195}{256}}\frac{1}{z^{16}}
\nonumber\\ &&
-{\frac{960627735}{256}}\frac{1}{z^{17}}
+{\frac {6381469401}{512}}\frac{1}{z^{18}}
+{\frac {15631424775}{128}}\frac{1}{z^{19}}
\Biggr] \ln(2)
\nonumber
\\
&&
+ \Biggl[
-{\frac {3}{4}}\frac{1}{z^{2}}
-{\frac {3}{8}}\frac{1}{z^{3}}
+{\frac {9}{16}}\frac{1}{z^{4}}
+{\frac {9}{16}}\frac{1}{z^{5}}
-{\frac {45}{32}}\frac{1}{z^{6}}
-{\frac {15}{8}}\frac{1}{z^{7}}
+{\frac {441}{64}}\frac{1}{z^{8}}
+{\frac {357}{32}}\frac{1}{z^{9}}
\nonumber\\ &&
-{\frac {3555}{64}}\frac{1}{z^{10}}
-{\frac {837}{8}}\frac{1}{z^{11}}
+{\frac {85437}{128}}\frac{1}{z^{12}}
+{\frac {22803}{16}}\frac{1}{z^{13}}
-{\frac {716625}{64}}\frac{1}{z^{14}}
-{\frac {212979}{8}}\frac{1}{z^{15}}
\nonumber
\end{eqnarray}\begin{eqnarray}
&&
+{\frac {64041849}{256}}\frac{1}{z^{16}}
+{\frac {41830605}{64}}\frac{1}{z^{17}}
-{\frac {919495575}{128}}\frac{1}{z^{18}}
-{\frac {163316841}{8}}\frac{1}{z^{19}}
\Biggr] \ln^2(2)
\nonumber
\\ &&
+ \Biggl[
{\frac {1}{2}}\frac{1}{z}
+{\frac {1}{4}}\frac{1}{z^{2}}
-{\frac {1}{8}}\frac{1}{z^{4}}
+{\frac {1}{4}}\frac{1}{z^{6}}
-{\frac {17}{16}}\frac{1}{z^{8}}
+{\frac {31}{4}}\frac{1}{z^{10}}
-{\frac {691}{8}}\frac{1}{z^{12}}
+{\frac {5461}{4}}\frac{1}{z^{14}}
\nonumber
\\
&&
-{\frac {929569}{32}}\frac{1}{z^{16}}
+{\frac {3202291}{4}}\frac{1}{z^{18}}
\Biggr] 
\ln^3(2) + O\left(\frac{1}{z^{20}}\right)~.
\end{eqnarray}
In the above
$T_k$ denotes the tangent numbers or {Euler}
zigzag numbers,~\cite{NIELS2,BERNOU}, with the generating function
\begin{eqnarray}
\frac{\exp(x)-\exp(-x)}{\exp(x)+\exp(-x)} = \sum_{k=1}^{\infty}
\frac{(-1)^{k-1}}{(2k-1)!} T_k x^{2k-1}~.
\end{eqnarray}
They are related to the { Bernoulli} numbers by
\begin{eqnarray}
T_k = \frac{2^{2k}\left(2^{2k}-1\right)}{2k} (-1)^{k-1}B_{2k}~.
\end{eqnarray}

\newpage
\section{\boldmath Some Integrals of $w = 5$}
%

\vspace{1mm}
\noindent
In this appendix a series of integrals used in the present calculation
is presented. Other integrals of this type were given in \cite{LRR}.
\begin{eqnarray} 
\int_0^1 dz \frac{\Li_4(z)}{1+z} 
&=&   \ln(2) \zeta_4 + \frac{3}{4} \zeta_2 \zeta_3 - \frac{59}{32} \zeta_5
\\
\int_0^1 dx \frac{\Li_4(x^2)}{1+x} &=& 
\frac{2}{5} \zeta_2^2 \ln(2) +3 \zeta_2
\zeta_3 - \frac{25}{4} \zeta_5 
\\
\int_0^1 dz \frac{S_{1,3}(z)}{1+z} 
&=& \ln(2) \zeta_4 - \frac{7}{16} \ZTW \ZTH-\frac{1}{6} \ZTW \ln^3(2) 
+ \frac{7}{16} \ZTH \ln^2(2) \nonumber\\ & &         
- \frac{27}{32} \ZFI +\ln(2) \lifhalf 
+ \frac{1}{30} \ln^5(2) + \liFhalf 
\\    
\int_0^1 dz \frac{S_{2,2}(z)}{1+z}   
&=& \frac{1}{4} \zeta_4 \ln(2) - \frac{7}{8} \ZTW \ZTH - \frac{1}{3} \ZTW \ln^3(2) 
+ \frac{7}{8}\ZTH \ln^2(2) \nonumber\\ & & 
- \frac{15}{32} \ZFI + 2 \left[ \ln(2) \lifhalf + \liFhalf \right] + \frac{1}{15} 
\ln^5(2) 
\\
\int_0^1 dz \frac{S_{2,2}(-z)}{1+z} &=&
-2 \liFhalf + \frac{125}{64} \zeta_5 - \frac{3}{4} \zeta_2^2 \ln(2) + \frac{7}{8} 
\zeta_3
\ln^2(2) \nonumber\\ & &
-\frac{1}{6} \zeta_2 \ln^3(2) + \frac{1}{60} \ln^5(2) 
\\
\int_0^1 dz \frac{\Li_2^2(z)}{1+z}   
&=&
\ln(2) \zeta_2^2- \frac{5}{4} \zeta_2 \zeta_3 + \frac{29}{32} \zeta_5 
\\
\int_0^1 dz \frac{\Li_2^2(-z)}{1+z}  &=& 
\frac{1}{4} \zeta_2^2 \ln(2) + \frac{7}{2} \zeta_3 \ln^2(2) -
\frac{4}{3} \zeta_2 \ln^3(2) - \frac{1}{4} \zeta_2 \zeta_3 \nonumber\\ & &
- \frac{125}{16} \zeta_5 + \frac{4}{15} \ln^5(2) + 8 \left[\Li_4\left(
\frac{1}{2}\right) \ln(2) + \Li_5\left(\frac{1}{2}\right)\right]
\\
\int_0^1 dx \frac{\ln(x) \Li_3(x)}{1+x} &=& \frac{21}{8} \zeta_2 \zeta_3 - 
\frac{83}{16} \zeta_5
\\
\int_0^1 dx \frac{\ln^3(x) \ln(1-x)}{1+x} &=& 
- \frac{9}{2} \zeta_2 \left[\zeta_3 + \zeta_2 \ln(2)\right] +
\frac{273}{16} \zeta_5 
\\
\int_0^1 dx \frac{\ln(x) \Li_3(-x)}{1+x} &=& 
\frac{51}{32} \zeta_5 - \frac{3}{4} \zeta_2 \zeta_3 \hspace{1cm} 
\\
\int_0^1 dx \Li_3(-x) \frac{\ln(1+x)}{1+x} &=& \frac{1}{16} \zeta_2
\zeta_3 + \frac{1}{3} \zeta_2 \ln^3(2) -\frac{5}{4} \zeta_3 \ln^2(2)
+ \frac{125}{64} \zeta_5
\nonumber\\
& &
- \frac{1}{15} \ln^5(2) - 2 \left[\Li_4\left(\frac{1}{2}\right) \ln(2) + 
 \Li_5\left(\frac{1}{2}\right)\right] 
\\
\int_0^1 \frac{dx}{x} \Li_2(-x) \ln^2(1+x) &=& \frac{7}{4} \zeta_3 \ln^2(2) 
- \frac{1}{8} \zeta_2 \zeta_3 - \frac{2}{3} \zeta_2 \ln^3(2) \nonumber\\
& & - \frac{125}{32} \zeta_5 + 4 \left[ \ln(2) \Li_4\left(\frac{1}{2}\right)
+ \Li_5\left(\frac{1}{2}\right)\right] +\frac{2}{15} \ln^5(2)
\\
\int_0^1 \frac{dx}{x} \Li_2(-x) \ln^2(1-x) &=& - \frac{7}{4} \zeta_3 \ln^2(2) 
+ \frac{3}{4} \zeta_2 \zeta_3 + \frac{2}{3} \zeta_2 \ln^3(2) \nonumber\\
& & + \frac{15}{16} \zeta_5 - 4 \left[ \ln(2) \Li_4\left(\frac{1}{2}\right)
+ \Li_5\left(\frac{1}{2}\right)\right] - \frac{2}{15} \ln^5(2)
\end{eqnarray}
\begin{eqnarray}
\int_0^1 dx \frac{\ln(x)}{1+x} S_{1,2}(x) &=& \frac{41}{32} \zeta_5 - \frac{11}{16} 
\zeta_2 \zeta_3 
\\
\int_0^1 \frac{dx}{x} \ln(x) \ln(1+x) \ln^2(1-x) &=& 
- \frac{7}{2} \zeta(5) - \frac{3}{8} \zeta(2) \zeta(3) + 4 \left[\Li_5\left(
\frac{1}{2} \right) + 4 \Li_4\left(\frac{1}{2}\right) \ln(2) \right] \nonumber
\\ & & 
+ \frac{7}{4} \zeta_3 \ln(2) - \frac{2}{3} \zeta_2 \ln^3(2) + \frac{2}{15} \ln^5(2)
\\
\int_0^1 \frac{dx}{x} \ln(x) \ln(1-x) \ln^2(1+x) &=& 
-\frac{25}{16}  \zeta_5 + \frac{7}{8} \zeta_2 \zeta_3
\\
\int_0^1~dx \frac{\Li_2(x^2)}{1+x} &=& -\frac{3}{4} \zeta_3 + \zeta_2 
\ln(2) 
\\ 
\int_0^1 dx~\frac{I_1(x)}{1+x} &=& - \frac{5}{8} \zeta_3 \ln(2)
+ \frac{3}{20} \zeta_2^2 
\\
\int_0^1 dx~\frac{\Sf(x^2)}{1+x} &=& 
\frac{9}{2} \zeta_3 \ln(2) - \frac{37}{20} \zeta_2^2
+ 4 \Li_4\left(\frac{1}{2}\right) - \zeta_2 \ln^2(2) + \frac{1}{6} \ln^4(2)
\nonumber\\
\end{eqnarray}

\newpage

\end{document}